\documentclass[12pt]{article}
\usepackage{amsmath}
\usepackage{graphicx}
\usepackage{enumerate}
\usepackage{natbib}
\usepackage{url} 
\usepackage{amsfonts}
\usepackage{amsmath}
\usepackage{amssymb}
\usepackage{amsthm}
\usepackage{enumerate}
\usepackage{graphicx}
\usepackage{layout}
\usepackage{latexsym}
\usepackage{array}
\usepackage{extarrows}
\usepackage{balance}
\usepackage{titlesec}
\usepackage{floatrow}  
\usepackage{rotating} 
\usepackage{color}
\usepackage{dcolumn}
\usepackage{multicol}
\usepackage{multicol}
\usepackage{setspace}
\usepackage{listings}
\usepackage{fancyhdr}
\usepackage{float}
\usepackage{graphicx}
\usepackage{subfigure}

\usepackage[linesnumbered,ruled,vlined]{algorithm2e}

\floatstyle{plaintop}
\restylefloat{table}
\usepackage[flushleft]{threeparttable}
\usepackage[CJKbookmarks=true]{hyperref}
\hypersetup{hidelinks}

\newtheorem{theorem}{Theorem}

\newtheorem{corollary}{Corollary}
\newtheorem{proposition}{Proposition}
\newtheorem{definition}{Definition}[section] 

\newtheorem*{remark}{Remark}
\newcommand{\blind}{1}

\newcommand{\bcon}{{\bf Condition }}

\newcommand{\btheta}{\boldsymbol{\theta}}

\newcommand{\bZ}{\mathbf{Z}}

\begin{document}

	\def\spacingset#1{\renewcommand{\baselinestretch}%
		{#1}\small\normalsize} \spacingset{1}

	
	\if1\blind
	{
		\title{\bf Perturbation-based Inference for Extreme Value Index}
		\author{Yiwei Tang\\
			Department of Statistics and Data Science, Fudan University\\
			and \\
			Judy Huixia Wang \\
			Department of Statistics, Rice University\\
			and \\
			Deyuan Li \\
			Department of Statistics and Data Science, Fudan University}
		\maketitle
	} \fi
	
	\if0\blind
	{
		\bigskip
		\bigskip
		\bigskip
		\begin{center}
			{\LARGE\bf Perturbation-based Inference for Extreme Value Index}
		\end{center}
		\medskip
	} \fi
	
	\bigskip
	\begin{abstract}
		The extreme value index (EVI) characterizes the tail behavior of a distribution and is crucial for extreme value theory. Inference on the EVI is challenging due to data scarcity in the tail region. We propose a novel method for constructing confidence intervals for the EVI using synthetic exceedances generated via perturbation. Rather than perturbing the entire sample, we add noise to exceedances above a high threshold and apply the generalized Pareto distribution (GPD) approximation. Confidence intervals are derived by simulating the distribution of pivotal statistics from the perturbed data. We show that the pivotal statistic is consistent, ensuring the proposed method provides consistent intervals for the EVI. Additionally, we demonstrate that the perturbed data is differentially private. When the GPD approximation is inadequate, we introduce a refined perturbation method. Simulation results show that our approach outperforms existing methods, providing robust and reliable inference.
		
	\end{abstract}
	
	\noindent%
	\vfill
	\newpage
	\spacingset{1.9} 

	\section{Introduction}
	\label{sec:intro}
	
	Extreme value theory (EVT) is concerned with the development of statistical models for inference on extreme values, either very large or very small. A commonly used model in this context is the generalized Pareto distribution (GPD), which describes exceedances over a high threshold through shape and scale parameters. This model underpins the peaks-over-threshold methodology introduced by Davison and Smith (1990). The GPD's shape parameter, known as the extreme value index (EVI), is central to EVT and plays a crucial role in the estimation of high quantiles and risk. 
	
	The point estimation for the EVI have been extensively studied \citep{de2007extreme,hill1975simple,pickands1975statistical}, while inference on the EVI has been less explored, with several approaches commonly used.  \cite{cheng1998asymptotic} and \cite{cheng2001confidence} constructed confidence intervals for the EVI of heavy-tailed distributions based on the limiting distribution of the EVI estimator, which is a standard method and known as normal approximation. Additionally, \cite{lu2002likelihood} employed both empirical likelihood and parametric likelihood methods to construct confidence intervals for the EVI. Bootstrap methods are also commonly applied in this context. \cite{guillou2000bootstrap} demonstrated that subsampling bootstrap methods generally perform less accurately than normal approximation but proposed a mixture of pivotal quantities to enhance their accuracy. In contrast, \cite{el2000bootstrap} validated the fullsample efficiency of the bootstrap method of a specific estimator of the EVI using the theoretical framework provided by \cite{chen1997mapping}. Recently, \cite{de2024bootstrapping} derived the tail quantile process of intermediate bootstrap samples, proving the consistency of the bootstrap method for a large family of EVI estimators.
	
	Two primary challenges in EVI inference are the lack of accurate finite-sample confidence intervals and the difficulty in selecting an appropriate threshold. First, methods such as normal approximation, empirical likelihood, and parametric likelihood all rely on asymptotic distribution theory. Bootstrap methods, based on Monte Carlo simulations, similarly require large sample sizes. Second, threshold selection is another major difficulty. While higher thresholds yield better approximations, they also increase variability in parameter estimates. Common graphical diagnostics \citep{drees2000make,coles2001introduction,scarrott2012review} are subjective and unsuitable for automation, motivating more recent quantitative approaches such as cross-validation \citep{northrop2017cross} and goodness-of-fit tests \citep{bader2018automated}.
	

	
	Data perturbation is a technique for generating synthetic data by adding "noise" to original data, with applications in various fields, including data security and privacy. In data privacy, perturbation can ensure a prescribed level of \textit{differential privacy}, a widely adopted framework that guarantees the output of an algorithm does not reveal much about the presence or absence of any single individual in the dataset. Common perturbation methods for achieving differential privacy include the Laplace and exponential mechanisms \citep{dwork2006differential}. In statistics and data science, data perturbation is used in model selection, selective inference, and synthetic generation. For example, \cite{ye1998measuring}, \cite{shen2002adaptive}, and \cite{wang2006estimation} used Gaussian perturbation to study the impact of model selection on estimation. In machine learning, \cite{rombach2022high}, \cite{dhariwal2021diffusion}, and \cite{sohl2015deep} applied Gaussian noise in denoising diffusion models to improve the diversity of generated samples. Recently, \cite{shen2022data} proposed a perturbation-based approach that generates synthetic data while preserving both privacy and predictive accuracy. By assuming and estimating a parametric distribution, this method allows for the generation of perturbed samples and subsequent statistical analyses, including pivotal inference. \cite{liu2024novel} extended this approach to unstructured data contexts, utilizing deep learning models.
	
	
	
	In this paper, we propose a method for constructing confidence intervals for the EVI using synthetic data generated through perturbation. Instead of adding Gaussian or Laplace noise to the original sample, we perturb exceedances above a high threshold by applying a nonlinear transformation to standardized random variables, which are then mapped to GPD random variables. Using the asymptotic pivotal statistic, we construct the confidence interval. Rather than relying on the limiting distribution of this pivotal statistic, we simulate its finite-sample distribution based on the perturbed pivotal statistic. We show that the pivotal statistic derived from the synthetic data is consistent, ensuring that the proposed confidence interval is also consistent, provided the GPD approximation is sufficiently accurate. Furthermore, we demonstrate that the perturbed data is differentially private, safeguarding sensitive information. In cases where the GPD approximation is not sufficiently accurate, we recommend a refined perturbation procedure using an extended GPD model, which offers greater flexibility by incorporating additional parameters to reduce bias in the EVI estimates. By combining perturbed pivotal statistics from both standard and refined GPD perturbations, we construct a confidence interval that provides accurate coverage and is more robust to threshold selection.
	
	The remainder of the paper is organized as follows. In Section 2, we outline the procedure for EVI inference using perturbation. Section 3 develops the theoretical results for the proposed confidence interval, including the tail quantile process of the perturbed sample, and establishes that the perturbed data satisfy differential privacy. Section 4 addresses computational issues, such as the selection of perturbation scales and strategies for dealing with the limitations of the GPD approximation. Finally, Section 5 presents simulation studies that assess the finite-sample performance of the proposed methods. Additional technical details are provided in the online Supplementary Material.
	
	
	\section{Perturbation-based Extreme Value Index Inference}
	\label{sec:ptb}
	In this section, we propose an inference procedure for the EVI based on perturbation. Since hypothesis testing and confidence interval estimation are closely related, in this paper we focus on the construction of confidence intervals. Suppose we observe a random sample $\{Y_1, Y_2,\ldots, Y_n\}$ from a distribution $F$. Throughout the article, we assume that \( F \) belongs to the maximum domain of attraction of an extreme value distribution \( H_\gamma(\cdot) \), denoted by \( F \in D(H_\gamma) \). This means, there exist constants \( a_n > 0 \) and \( b_n \in \mathbb{R} \) such that
	\[
	P\left(\frac{\max_{1 \leq i \leq n} Y_i - b_n}{a_n} \leq y\right) \to H_\gamma(y) := \exp \left\{-(1 + \gamma y)^{-1 / \gamma}\right\},
	\]
	as \( n \to \infty \), for \( 1 + \gamma y \geq 0 \), where $\gamma$ is the EVI. In this paper, we focus on the case $\gamma>0$, which corresponds to heavy-tailed distributions, as commonly encountered in finance (e.g., stock returns), insurance (e.g., large claims), seismology (e.g., earthquake magnitudes), and hydrology (e.g., river flows during floods). Given the original sample \( \{Y_1, Y_2, \ldots, Y_n\} \), we select the largest \( k+1 \) observations, \( Y_{n-k,n} \le Y_{n-k+1,n}\le \ldots \le Y_{n,n} \), and define \( Z_i = Y_{n-i+1,n} - Y_{n-k,n} \) for \( i = 1, 2, \ldots, k \). Typically, we choose \( k = k(n) \) such that \( k \to \infty \) and \( k/n \to 0 \) as \( n \to \infty \), which is known as an intermediate sequence in EVT. Define the cumulative distribution function (cdf) of the GPD as
	\[
	G_{\gamma, \beta}(y) =\left\{
	\begin{aligned}
		1 - (1 + \gamma y / \beta)^{-1 / \gamma},\quad & \gamma \neq 0, \\
		1 - \exp(-y / \beta),\quad & \gamma = 0,
	\end{aligned}
	\right.
	\]
	for $y\in(0,(0 \vee(-\gamma))^{-1})$, where $0^{-1}$ is defined to be $\infty$, and \( \gamma \) and \( \beta>0 \) denote the shape and scale parameters, respectively.

	The following Pickands-Balkema-de Haan theorem \citep{pickands1975statistical} provided an equivalent condition for \( F \in D(H_\gamma) \).
	
	\textbf{Pickands-Balkema-de Haan Theorem.  }\( F \in D(H_\gamma) \) if and only if there exists a positive function \( \beta(u) \) such that
	\[
	\lim_{u \rightarrow y_F} \sup_{0 \leq y < y_F - u} \left| \frac{\bar{F}(u + y)}{\bar{F}(u)} - \bar{G}_{\gamma, \beta(u)}(y) \right| = 0,
	\]
	where \( y_F = \sup \{ s : F(s) < 1 \} \), \( \bar{F} := 1 - F \), and $\bar{G}_{\gamma, \beta(u)}=1-{G}_{\gamma, \beta(u)}$.
	
	The Pickands-Balkema-de Haan theorem implies that for large \( t \), \( Y - t \), conditional on \( Y > t \), approximately follows a GPD with parameter \( \btheta = (\gamma, \beta) \). Based on this asymptotic distribution, we perturb the tail exceedance sample \( \{Z_1, Z_2, \ldots, Z_k\} \). Suppose we obtain an estimator of \( \btheta \), denoted as \( \hat{\btheta} = (\hat{\gamma}, \hat{\beta}) \). We apply the following perturbation-based scheme to generate synthetic data that follows the GPD with parameter \( \hat{\btheta} \). First, for each $i = 1, 2, \ldots, m$, independently generate $k$ centralized Laplace random variables $\{e_{ij}, j = 1, 2, \ldots, k\}$ with scale parameter $b>0$. That is, each $e_{ij}$ follows the Laplace distribution with density function $f_e(x) = {b}/{2} \exp(-b|x|), x\in\mathbb{R}$, denoted by $e_{ij}\sim$Laplace$(0, b)$. Second, generate \( Z_{ij}^* \) by $Z_{ij}^* = H\left(G_{\hat{\btheta}}(Z_j)+ e_{ij}\right), i = 1, 2, \ldots, m,  j = 1, 2, \ldots, k$, where \( G_{\hat{\btheta}} \) is the cdf of the GPD with parameter \( \hat{\btheta} \), and  \( H(\cdot) = G_{\hat{\btheta}}^{-1}\left(R(\cdot)\right) \). Here, \( R \) denotes the cdf of $U+e_{ij}$, where $U\sim U(0,1)$ independent of $e_{ij}$. Given the $m$ perturbed exceedance samples \( \{Z_{ij}^*,  j = 1, 2, \ldots, k\} \),  $i = 1, 2, \ldots, m$, we obtain the \( i \)-th perturbed sample \( \mathbf{Y}_i^*= \{Y_{1,n}, Y_{2,n}, \ldots, Y_{n-k,n}, Y_{n-k,n} + Z_{i1}^*, Y_{n-k,n} + Z_{i2}^*, \ldots, Y_{n-k,n} + Z_{ik}^*\} \) for \( i = 1, 2, \ldots, m \). Notice that $Z_{ij}^*$ are all positive value because they are drawn from GPD, and hence the largest $k+1$ observations of the perturbed sample are $Y_{n-k,n}+Z^*_{i1},  Y_{n-k,n}+Z^*_{i2}, \ldots,  Y_{n-k,n}+Z^*_{ik}$. Without loss of generality, we assume $Z^*_{i1}\le Z^*_{i2}\le \cdots\le Z^*_{ik}$.

	Conditional on the original sample $\{Y_1,Y_2,\ldots, Y_n\}$, $Z_{ij}^*$ asymptotically preserves the distribution of $G_{\hat{\gamma}, \hat{\beta}}$, while ensuring differential privacy. 
	The perturbation approach offers several advantages. First, instead of directly adding noise to $Z_j$, which would distort the perturbed sample’s distribution, we add noise to a standardized random variable $G_{\hat{\btheta}}(Z_j)$ and then transform the sample to match the GPD. Second, while applying the Laplace mechanism to heavy-tailed and unbounded data is impractical due to the challenge of finding a finite scale, normalization allows for the effective use of the Laplace mechanism to preserve privacy.

	We construct the confidence interval for $\gamma$ through pivotal inference. Focusing on heavy-tailed distributions, we utilize the Hill estimator \citep{hill1975simple}, defined as $\hat{\gamma}=k^{-1}\sum_{i=0}^{k-1}$ $\log\left({Y_{n-i,n}}/{Y_{n-k,n}}\right)$. 
	We construct an asymptotic pivotal given by $T\left(\gamma,\hat{\gamma}\right)=\sqrt{k} (\hat{\gamma}-\gamma)/\hat{\gamma}$,
	which asymptotically follows $N\left(0,1\right)$ under certain conditions and is also applied by \cite{cheng2001confidence}. The perturbed version of $T$ is constructed as $T^*=T\left(\hat{\gamma},\hat{\gamma}^*\right)$ where $\hat{\gamma}^*$ is the Hill estimator based on the perturbed sample. For $i=1,2,\ldots,m$, we obtain $\hat{\gamma}_{i}^*$ using $\mathbf{Y}_i^*$ and compute the perturbed pivotal by $T_i^* = T\left(\hat{\gamma},\hat{\gamma}_{i}^*\right)$. We approximate the distribution of $T$ by the empirical distribution of $T^*$, $\{T_i^*, i=1,2,\ldots,m\}$, and construct the confidence interval using sample percentiles. Denote the $\alpha$th sample quantile of $T_{i}^*$s as $Q_{T^*}\left(\alpha\right)$. To clarify the procedure, we summarize it in Algorithm 1.

	
	The choice of $\hat{\btheta}$ is crucial in the perturbation procedure. Note that the true parameter $\beta$ is not unique and depends on the threshold. According to Theorem 1.2.5 in \cite{de2007extreme}, when $\gamma>0$, the GPD approximation holds if the scale parameter \(\beta = t\gamma\), where \(t\) is the threshold. Therefore, we consider \(\beta=\gamma U(n/k)\), where  $U(t):=\inf \{y: F(y)\ge 1-1/t\}$ is the $(1-1/t)$-quantile of $Y$, and estimate $\beta$ by \(\hat{\gamma} Y_{n-k,n}\), leading to the estimator \(\hat{\btheta} = \left(\hat{\gamma}, \hat{\gamma} Y_{n-k,n}\right)\).


	There are three key differences between our approach and that of \cite{shen2022data}. First, while \cite{shen2022data} assumed a specific parametric distribution for the original sample, our method relies on a maximum domain of attraction condition, which applies to a wider range of distributions. Second, \cite{shen2022data} perturbed the entire original sample based on the parametric assumption, whereas our approach perturbs only the exceedances above a threshold, using their asymptotic distribution as the threshold tends to infinity. Therefore, we construct asymptotic pivotal and it adds complexity to the derivation of theoretical results. Third, \cite{shen2022data} assumed the data-generating distribution is fully known, while our method handles the case where the distribution is unknown, and estimates the tail distribution from the data, introducing estimation error and complicating the analysis. Consequently, our procedure requires careful selection of the perturbation scale to preserve the tail behavior.
	
	

	\begin{algorithm}
		\SetKwInOut{Input}{Input}
		\SetKwInOut{Output}{Output}
		\SetKwProg{Fn}{Algorithm}{}{\KwRet end}
		
		\caption{Confidence Interval for EVI Based on Data Perturbation}
		\Input{Data $\mathbf{Y} = \{Y_i\}_{i=1,2,\ldots,n}$, tail size $k$, perturbation scale $b$, confidence level $\alpha$}
		\Output{Confidence interval $I_{\text{ptb}}$}
		
		\Fn{PtbPivotal$(\mathbf{Y}, k, b)$}{
			Calculate the Hill estimator $\hat{\gamma}$ using $Y_{n-k,n} \le \ldots \le Y_{n,n}$\;
			\For{$i = 1$ \KwTo $m$}{
				\For{$j = 1$ \KwTo $k$}{
					Generate $e_{ij}$ independently from Laplace$(0, b)$\;
					Generate $Z_{ij}^*$ by $Z_{ij}^* = H(G_{\hat{\btheta}}(Z_j) + e_{ij})$\;
				}
				Calculate the perturbed Hill estimator
				\[
				\hat{\gamma}_{i}^* = \frac{1}{k}\sum_{j=1}^{k} \log\left(\frac{Z_{ij}^* + Y_{n-k,n}}{Y_{n-k,n}}\right)
				\]
				Calculate the perturbed pivotal as $T_i^* = T(\hat{\gamma}, \hat{\gamma}_{i}^*)$\;
			}
			\Return $\{T_i^*, i=1, 2, \ldots, m\}$ and $\hat{\gamma}$\;
		}
		
		\Fn{CIPivotal($\{T_i^*\}_{i=1}^m$, $\hat{\gamma}$)}{
			\Return $I_{\text{ptb}} = \left[ \left( 1 - \frac{1}{\sqrt{k}} Q_{T^*}\left( 1 - \frac{\alpha}{2} \right) \right) \hat{\gamma}, \; \left( 1 - \frac{1}{\sqrt{k}} Q_{T^*}\left( \frac{\alpha}{2} \right) \right) \hat{\gamma} \right]$\;
		}
		
		Compute $(\{T_i^*\}_{i=1}^m, \hat{\gamma}) \gets$ \textit{PtbPivotal}$(\mathbf{Y}, k, b)$\;
		Obtain $I_{\text{ptb}} \gets$ \textit{CIPivotal}($\{T_i^*\}_{i=1}^m$, $\hat{\gamma}$).
	\end{algorithm}

	\section{Theoretical Results}
	\label{sec:theo}

	In this section, we establish the consistency of the perturbed sample with respect to the pivotal quantity. Specifically, we first derive the perturbation analogue of the tail quantile process for the original sample (Theorem \ref{theorem:tqp}), and then analyze the asymptotic behavior of the perturbed Hill estimator (Corollary \ref{cor1}). Under the assumption of negligible bias (i.e., $\sqrt{k}A(n/k)\to 0$; see below for details), we show that the perturbed sample is consistent for the pivotal quantity (Theorem \ref{theorem:consistency}), where consistency is defined analogously to the bootstrap consistency of Bickel and Freedman (1981). Finally, we demonstrate that the perturbation procedure ensures differential privacy when $\btheta$ is known, while a mild condition is required to guarantee differential privacy when $\btheta$ is unknown.

	Before presenting the main conclusions of this paper, we first review some key existing results. 
	

	
	\bcon 1. Assume there exist \(\rho \leq 0\), a positive function \(a\), and a function \(A\) that is eventually positive or negative with \(\lim_{t \to \infty} A(t) = 0\), such that for \(x > 0\),
	
	$$
	\lim_{t \to \infty} \frac{\frac{U(t x) - U(t)}{a(t)} - \frac{x^\gamma - 1}{\gamma}}{A(t)} = \Psi_{\gamma, \rho}(x),
	$$where
	$$
	\Psi_{\gamma, \rho}(x) = \begin{cases}
		\frac{x^{\gamma + \rho} - 1}{\gamma + \rho}, & \text{if } \rho < 0, \\
		\frac{1}{\gamma} x^\gamma \log x, & \text{if } \rho = 0 \text{ and } \gamma \neq 0, \\
		\frac{1}{2} (\log x)^2, & \text{if } \rho = 0 \text{ and } \gamma = 0.
	\end{cases}
	$$

	Here, $A(\cdot)$ is regularly varying at infinity with index $\rho\le 0$, denoted by $A\in RV(\rho)$. The parameters \(\gamma\) and \(\rho\) are referred to as the EVI and the second-order index, respectively. Condition 1 is also known as the second-order condition in EVT.
	
	Below, we present a transformed version of the tail quantile process, $\{Y_{n-[ks],n}, s\in [0,1]\}$, for heavy-tailed distributions, which simplifies Theorem 2.1 in \cite{drees1998smooth}, as well as Theorems 2.4.2 and 2.4.8 in \cite{de2007extreme}.
	\begin{proposition}
		\label{Prop:tqp}
		Assume Condition 1 holds for some $\gamma>0$ and $\rho\le 0$. Let $k=k(n)\rightarrow\infty$ and $k/n\rightarrow 0$ as $n\rightarrow\infty$ and $\sqrt{k}A\left({n}/{k}\right)=O\left(1\right)$ as $n\rightarrow\infty$. There exist suitable axulliary function $A_0$ and a sequence of Brownian Motions $\left\{W_n\left(s\right)\right\}_{s\ge 0}$ such that, for any $\epsilon>0$, as $n\rightarrow\infty$, 
		$$
		\sqrt{k}\left(\frac{Y_{n-[k s], n}}{U\left(\frac{n}{k}\right)}-s^{-\gamma}\right)=\gamma s^{-\gamma-1} W_n(s)+\sqrt{k} A_0\left(\frac{n}{k}\right) s^{-\gamma} \frac{s^{-\rho}-1}{\rho} +s^{-\gamma-1 / 2-\epsilon} o_p\left(1\right),
		$$
		$$\sqrt{k}\left(\frac{\log Y_{n-\left[ks\right],n}-\log U\left(\frac{n}{k}\right)}{\gamma}+\log s\right)=s^{-1}W_n\left(s\right)+\sqrt{k}A_0\left(\frac{n}{k}\right)\frac{1}{\gamma}\frac{s^{-\rho}-1}{\rho}+s^{-\frac{1}{2}-\epsilon}o_p\left(1\right),$$
		uniformly for all $s\in\left(0,1\right]$.
	\end{proposition}
	


		We next review the classical results related to the Hill estimator in the context of the tail quantile process, as detailed in \cite{de2007extreme}.
		
		\begin{proposition}
			\label{Prop:hill}
			Assume the conditions of Proposition \ref{Prop:tqp} hold. Then, we have
			\begin{equation}
				\label{hill}
				\sqrt{k}\left(\hat{\gamma}-\gamma\right)=\gamma\int_{0}^{1}\left(s^{-1}W_n\left(s\right)-W_n\left(1\right)\right) ds+\sqrt{k}A_0\left(\frac{n}{k}\right)\frac{1}{1-\rho}+o_p\left(1\right).
			\end{equation}
		\end{proposition}
		Note that the term \(\int_{0}^{1} \left(s^{-1} W_n(s) - W_n(1)\right) ds\) is normal distributed with zero mean and variance equal to one by the definition of Brownian motion. Consequently, assuming \(\sqrt{k} A_0\left({n}/{k}\right)\rightarrow\lambda\) , we have $\sqrt{k}\left(\hat{\gamma} - \gamma\right) \overset{d}{\rightarrow} N\left({\lambda}/{(1 - \rho)}, \gamma^2\right)$. Specially, if \(\lambda=0\), \(\hat{\gamma}\) is asymptotically unbiased.

		We focus on the perturbed samples $\mathbf{Y}_i^*, i=1,2,\ldots,m$. Since they are identically distributed, we treat them as a generic sample $\mathbf{Y}^*$ and omit the index $i$ in theoretical results. The order statistics are denoted as $Y^*_{1,n} \leq Y^*_{2,n} \leq \cdots \leq Y^*_{n,n}$, and similarly, we write $T_i^*$ and $\hat{\gamma}_i^*$ as $T^*$ and $\hat{\gamma}^*$, respectively.


		\subsection{Consistent Confidence Interval}
		We first derive the tail quantile process of $\mathbf{Y}^*$ constructed in Algorithm 1.
		
		\begin{theorem}
			\label{theorem:tqp}
			Assume Condition 1 holds for some $\gamma>0$ and $\rho\le 0$. Let $k\left(n\right)\rightarrow\infty$, $b<(k^{-1}(k/n)^\rho) \wedge (k^{-1/2})$, and $\sqrt{k}A\left({n}/{k}\right)=O\left(1\right)$ as $n\rightarrow\infty$. Then there exists a sequence of Brownian motions $\left\{W^*_k\left(s\right)\right\}_{s \ge 0}$, independent of $\left\{Y_j\right\}_{j=1}^{n}$ and $\left\{W_n\left(s\right)\right\}_{s \ge 0}$ (defined in Proposition \ref{Prop:tqp}), such that for any $\epsilon>0$, as $n\rightarrow\infty$, 
			$$\begin{aligned}
				&\sqrt{k}\left(\frac{Y^*_{n-[k s], n}}{U\left(\frac{n}{k}\right)}-s^{-\gamma}\right)\\
				=&\gamma s^{-\gamma}\Big[s^{-1}W_k^*\left(s\right)-W_k^*\left(1\right)+\left.W_n\left(1\right)-\log\left(s\right)\int_{0}^{1}\left(s^{-1}W_n\left(s\right)-W_n\left(1\right)\right) ds\right. \\
				&-\log\left(s\right)\gamma^{-1}\sqrt{k}A_0\left(\frac{n}{k}\right)\frac{1}{1-\rho}\Big]+s^{-\gamma-1/2-\epsilon}\left(1-s\right)^{-\epsilon-1/2}o_p\left(1\right),
			\end{aligned}$$
			where $A_0$ is the same as in Proposition \ref{Prop:tqp}, and that 
			$$\begin{aligned}
				&\sqrt{k}\left(\frac{\log Y^*_{n-\left[ks\right],n}-\log U\left(\frac{n}{k}\right)}{\gamma}+\log s\right)\\
				=&s^{-1}W^*_k\left(s\right)-W^*_k\left(1\right)+W_n\left(1\right)-\log\left(s\right)\int_{0}^{1}\left(s^{-1}W_n\left(s\right)-W_n\left(1\right)\right) ds
				\\&-\log\left(s\right)\gamma^{-1}\sqrt{k}A_0\left(\frac{n}{k}\right)\frac{1}{1-\rho}+s^{-1/2-\epsilon}\left(1-s\right)^{-\epsilon-\frac{1}{2}}o_p\left(1\right),
			\end{aligned}$$
			uniformly for all $s\in\left(0,\frac{k}{k+1}\right]$.
		\end{theorem}
		\begin{remark}
			\label{remark1}
			The randomness in the tail quantile process $\{ Y_{n-\lfloor ks \rfloor, n}^*,\, s \in (0, k/(k+1)] \}$ arises from two sources. The first is introduced by the Brownian bridge $s^{-1} W_k^*(s) - W_k^*(1)$, which results from the perturbation procedure and the Generalized Pareto Distribution (GPD) approximation. The second source comes from the Brownian motion $W_n$, which reflects the estimation error in the original estimator. According to Propositions \ref{Prop:tqp} and \ref{Prop:hill}, the randomness in the scale and Extreme Value Index (EVI) estimators is driven by $W_n(1)$ and the integral  $\int_0^1 \left( s^{-1} W_n(s) - W_n(1) \right) ds$, which together capture the randomness in $Y^*$ due to the original sample.
			
			We compare the randomness arising from the GPD approximation in Theorem \ref{theorem:tqp} with the results in Proposition \ref{Prop:tqp}. First, the randomness in the original sample is driven by the Brownian motion $W_n$, rather than the Brownian bridge. Second, our results apply over a different uniform range for $s$. These differences arise because we perturb only the largest $k$ observations, introducing distinct randomness that complicates establishing a uniform bound for the interval $(k/(k+1), 1]$ for all $k$. Consequently, our results cannot be directly extended to the entire interval $(0,1]$, as in Proposition \ref{Prop:tqp}. Third, the second-order term in the tail quantile process of the original sample is absent in the perturbed version due to the GPD approximation, which effectively neglects second- and higher-order terms.

			The perturbation scale condition $b < \min \left( k^{-1} (k/n)^\rho, k^{-1/2} \right)$ ensures that the perturbation scale $b=b(n)$ decays sufficiently fast, preserving $R(G_{\hat{\btheta}}(Z_j) + e_{ij})$ with the approximately same distribution as the uniform distribution. Importantly, the error between $G_{\hat{\btheta}}(Z_j)$ and the uniform distribution includes both higher-order biases and estimation errors. The condition $b < \min \left( k^{-1} (k/n)^\rho, k^{-1/2} \right)$ guarantees that the scale of $e_{ij}$ is large enough for the second-order approximation error to be negligible and ensures the estimation error is controlled, so that $R(G_{\hat{\btheta}}(Z_j) + e_{ij})$ remains approximately uniform. In practice, $b$ should be tuned carefully to preserve information from the original sample. A more detailed discussion of choosing $b$ can be found in Section \ref{sec:comp}.

		\end{remark}
		Based on Theorem \ref{theorem:tqp}, the asymptotic disribution of the perturbed Hill estimator is obtained, showing that it is unbiased around the original Hill estimator.
		\begin{corollary}
			\label{cor1}
			Under the conditions of Theorem \ref{theorem:tqp}, we have
			$$\begin{aligned}
				\sqrt{k}\left(\hat{\gamma}^*-\gamma\right)=&\gamma\int_{0}^{1}\left(s^{-1}W^*_k\left(s\right)-W^*_k\left(1\right)\right) ds+\gamma\int_{0}^{1}\left(s^{-1}W_n\left(s\right)-W_n\left(1\right)\right) ds\\
				&+\sqrt{k}A_0\left(\frac{n}{k}\right)\frac{1}{1-\rho}+o_p\left(1\right).
			\end{aligned}$$
			Combining with result of Proposition \ref{hill}, we have
			\begin{equation*}
				\label{ptbhill}
				\sqrt{k}\left(\hat{\gamma}^*-\hat{\gamma}\right)=\gamma\int_{0}^{1}\left(s^{-1}W^*_k\left(s\right)-W^*_k\left(1\right)\right) ds+o_p\left(1\right).
			\end{equation*}
		\end{corollary}
		\begin{remark}
			According to Theorem 1 in \cite{shen2022data}, $T^* | \{Y_1, Y_2, \ldots, Y_n\}$ and $T$ have the same finite sample distributions. However, in this context, we only establish asymptotic equivalence in distribution, as $T$ is an asymptotic pivotal statistic rather than an exact pivotal statistic. While the finite-sample recovery properties of the perturbation procedure are somewhat compromised by its dependence on an exact pivotal statistic, the simulation results presented in Section 5 demonstrate that the proposed methods still outperform bootstrap methods in finite samples.
			
			A similar outcome is discussed in Section 4 of \cite{de2024bootstrapping} for the probability-weighted moment estimator using bootstrap samples. In this case, the absence of additional randomness related to the second-order condition results from the bootstrap procedure, which relies on the original sample.
		\end{remark}
		
	Note that the terms \(\gamma\int_{0}^{1}\left(s^{-1}W_n\left(s\right) - W_n\left(1\right)\right) \, ds\) and \(\gamma\int_{0}^{1}\left(s^{-1}W^*_k\left(s\right) - W^*_k\left(1\right)\right) \, ds\) are identically and independently distributed (i.i.d.). Comparing the results, $\hat{\gamma}-\gamma$ and $\hat{\gamma}^*-\hat{\gamma}$ are asymptotically identically distributed if the asymptotic bias term in $\hat{\gamma}-\gamma$ diminishes. Therefore, as demonstrated in the following theorem, if the bias of the original estimator is negligible, consistency results can be obtained.
		\begin{theorem}
			\label{theorem:consistency}
			Assume the conditions in Theorem \ref{theorem:tqp}. Assume $\sqrt{k}A\left({n}/{k}\right)=o\left(1\right)$ as $n\rightarrow \infty$. Then $T^*$ in Algorithm 1 is consistent, which means that as $n\rightarrow\infty$,
			$$\underset{x\in\mathbb{R}}{\sup}\left|P\left(T^*\le x| Y_1,Y_2,\ldots ,Y_n\right)-P\left(T\le x\right)\right|\overset{p}{\rightarrow} 0.$$
		\end{theorem}

		\subsection{Differential Privacy}

		Differential privacy ensures that replacing a single observation in a dataset has only a minimal impact on the released information, as quantified by $\varepsilon$-differential privacy. Suppose $\mathbf{Z}$ is a random sample from a cumulative distribution function $F_Z$, and $\mathfrak{m}$ is a privatization mechanism that maps $\mathbf{Z}$ to a public release $\tilde{\mathbf{Z}}$. Let $\mathbf{z}$ and $\mathbf{z}'$ be two adjacent realizations of $\mathbf{Z}$ differing in exactly one observation. Then, $\varepsilon$-differential privacy is defined as follows.
		
		\begin{definition}[\cite{dwork2006differential}]
			A privatization mechanism $\mathfrak{m}(\cdot)$ satisfies $\varepsilon$-differential privacy if
			$$
			\sup_{\mathbf{z}, \mathbf{z}'} \sup_B 
			\frac{P(\mathfrak{m}(\mathbf{Z}) \in B \mid \mathbf{Z}=\mathbf{z})}
			{P(\mathfrak{m}(\mathbf{Z}) \in B \mid \mathbf{Z}=\mathbf{z}')} 
			\le e^{\varepsilon},
			$$
			where $B$ is a measurable set and $\varepsilon \ge 0$ is a small parameter called the privacy budget. The ratio is interpreted to be 1 whenever the numerator and denominator are both 0.
		\end{definition}
		
		Heavy-tailed data pose unique challenges for privacy preservation. Because extreme values occur with relatively high probability, modifying a single tail observation can drastically alter the released information. Intuitively, achieving differential privacy in this setting may seem to require adding unbounded noise. Our method overcomes this issue by injecting bounded noise into a standardized variable and then transforming it back to the original scale. We now establish the privacy-preserving properties of the proposed mechanism, beginning with the case where the parameter $\boldsymbol{\theta}$ is known.

		\begin{theorem}
			\label{theorem:privacy}
			Assume the parameter $\boldsymbol{\theta}$ is known. For each $i = 1, 2, \ldots, m$, the mechanism for generating $\bZ_i^* = \{Z_{ij}^*, j = 1, 2, \ldots, k\}$ in Algorithm 1 is $b$-differentially private.
		\end{theorem}
		
		In practice, the parameter $\boldsymbol{\theta}$ is often unknown. To preserve differential privacy in this case, we suggest estimating the parameter from an independent sample that will not be privatized.
		
		\begin{theorem}
			\label{theorem:privacy1}
			Assume the parameter $\boldsymbol{\theta}$ is estimated from the sample $\tilde{\bZ}=\{\tilde{Z_1},\tilde{Z_2},\ldots, \tilde{Z_k}\}$, which is independent of $\{Z_1,Z_2,\ldots, Z_k\}$, denoted by $\hat{\btheta}(\tilde{\bZ})$. For each $i = 1, 2, \ldots, m$, the mechanism for generating $\bZ_i^* = \{Z_{ij}^*, j = 1, 2, \ldots, k\}$ in Algorithm 1 is $b$-differentially private.
		\end{theorem}
		
		In practice, obtaining an independent sample can be challenging. According to \cite{bi2023distribution}, one approach is to estimate the parameter using a random subsample of $\mathbf{Z}$, while privatizing the remaining data. Another option is to use a public dataset drawn from the same distribution.


		\section{Computational Issues}
		\label{sec:comp}
		The proposed procedure depends on the choice of the perturbation scale. Additionally, when the bias in the GPD approximation is significant, the proposed methods can be further refined. In this section, we discuss these practical computational issues.
		\subsection{Sensitivity Analysis}
		In this subsection, we perform a sensitivity analysis of the confidence interval with respect to the perturbation scale $b$. To investigate how varying perturbation scales affect inference, we conduct experiments with different distributions and perturbation scales, and assess the behavior of the corresponding confidence intervals.
		
		According to \cite{bi2023distribution} and Theorem \ref{theorem:privacy}, the perturbation scale is closely related to the privacy budget. However, in our context, we simply focus on the impact of $b$ on the finite distribution of $T^*$ or equivalently the finite distribution of $Z^*_{ij}$, which is crucial for a valid inference of the EVI. As \(b\) decreases to zero, the scale of the Laplace noise \(e_{ij}\) grows and eventually dominates \(G_{\hat{\btheta}}(Z_j)\), making the perturbed sample approximately follows $G_{\hat{\btheta}}$. Conversely, as \(b\) increases, \(G_{\hat{\btheta}}(Z_j)\) increasingly dominates \(e_{ij}\), causing \(R\) to converge to the cdf of the uniform distribution and \(Z_{ij}^*\) to approach \(Z_j\). While this preserves key information from the original sample, such as the rank of $Z_j$, it reduces the perturbation effect and limits the functional variability introduced by noise. Therefore, selecting an appropriate perturbation scale is essential. Motivated by Theorem \ref{theorem:tqp}, we choose the largest $b$ satisfying $b<(k^{-1}(k/n)^\rho) \wedge (k^{-1/2})$ to balance the effectiveness of the perturbed sample and the information from original sample. We choose $b=c_1 k^{-0.51}$ with \(c_1\in[1.5,3.5]\) and the results in Supplementary Material show that the performance of the confidence interval remains stable for  \(c_1\in[1.5,3.5]\), across various scenarios, sample sizes, and distributions. Throughout our numerical studies in Section \ref{sec:simul}, we use  \(c_1=2.5\).
		
		\subsection{Refined Generalized Pareto Distribution}
		\label{subsec:RGPD}
		Recall from Section \ref{sec:theo} that to ensure the consistency of the perturbed confidence interval, we assume that the asymptotic bias of the EVI vanishes, i.e., \(\sqrt{k} A\left({n}/{k}\right) = o(1)\). This condition implies that the tail sample size $k$ should be much smaller than the total sample size $n$. Since $A\in RV(\rho)$, the index \(\rho\) governs the rate at which \(A(t)\to0\). A small \(|\rho|\) may indicate a non-negligible bias. In practice, we observe that if the asymptotic bias is substantial, the perturbed pivotal \(T^*\) constructed in Algorithm 1 may deviate from the true pivotal \(T\), particularly when the threshold is relatively low. This observation suggests that the GPD may not be an appropriate model for capturing the distribution of excesses in such scenarios.

		To model excesses over thresholds, several flexible approaches have been proposed, as discussed in Section \ref{sec:intro}. Among these, \cite{beirlant2009second} proposed using a refined generalized Pareto distribution (RGPD) model for heavy-tailed distributions to capture exceedances. This model is particularly appealing because it strikes a balance between flexibility and variance by introducing an additional parameter. This parameter has a direct connection to the second-order coefficient \(\rho\) in Condition 1 (see \cite{beirlant2009second}), providing an interpolation that enhances the model's applicability. Specifically, the RGPD with parameter $\left(\gamma, \delta, \tau\right)$ is defined by its distribution function:
		$$
		{H}_{\gamma, \delta, \tau}(y)= \begin{cases}1-\left\{\left(1+y\right)\left(1+\delta-\delta \left(1+y\right)^\tau\right)\right\}^{-1 / \gamma}, & \text { if } y>0, \\ 0, & \text { if } y \leqslant 0.\end{cases}
		$$
		The GPD with positive shape parameter $\gamma>0$ and scale parameter $\beta>0$ is a member of the RGPD family by taking $\tau=-1$ and $\delta=\gamma/\beta-1$. The parameter $\left(\gamma, \delta, \tau\right)$ are estimated using the RGPD likelihood function, as detailed in \cite{beirlant2009second}, based on the $k$ exceedances ${Y_{n-i,n}}/{Y_{n-k,n}}, i=0,1,\ldots,k-1$. These estimators are denoted as $\left(\hat{\gamma}_{d}, \hat{\delta}, \hat{\tau}\right)$. Note that $\hat{\gamma}_{d}$ differs from the original estimator $\hat{\gamma}$.
		
		Given the estimator $\left(\hat{\gamma}_{d}, \hat{\delta}, \hat{\tau}\right)$, we replace the GPD approximation in Algorithm 1 with the RGPD to generate a new perturbed sample. Specifically, we compute $Z_{ij}^*$ by numerically solving the following equation:
		$$G_{\hat{\btheta}}(Z_j)+e_{ij}={H}_{\hat{\gamma}_{d}, \hat{\delta}, \hat{\tau}}(Z_{ij}^*/ Y_{n-k,n}),\qquad i=1,2,\ldots,m,\ldots j=1,2,\ldots, k.$$
		For $i=1,2,\ldots,m$, we apply the Hill method to the data $\left(Z_{i1}^*,Z_{i2}^*,\ldots,Z_{ik}^*\right)$ to obtain $\hat{\gamma}_{i}^*$.
		We then compute the perturbed pivotal statistic:
		$T_{i,d}^*=T\left(\hat{\gamma}_{d},\hat{\gamma}_{i}^*\right)$, for $ i=1,2,\ldots,m$. To approximate the distribution of $T$, we can use the empirical distribution of 
		$\{T_{i,d}^*, i=1,2,\ldots,m\}$. However, in this study, we simulate the distribution of $T$ through a mixture of RGPD and GPD approximations $T_w=wT_{GPD}+(1-w)T_{RGPD}$,
		where $w\in (0,1)$, $T_{GPD}$ and $T_{RGPD}$ are the pivotal statistics based on the GPD and RGPD approximations, respectively. Confidence intervals are then constructed using the distribution of the perturbed $T_w$. The rationale behind this approach is that while $T_{GPD}$ is biased but has low variance, $T_{RGPD}$ is unbiased but has high variance. The idea of weighted pivotal is inspired by \cite{guillou2000bootstrap}, which combined the pivotal of the subsample bootstrap with normal approximation to construct confidence intervals that perform better than either method alone.
		
		We set $w$ by $w={{1}/{W^2_{GPD}}}/{({1}/{W^2_{GPD}}+{1}/{W^2_{RGPD}})}$, where $W_{GPD}$ and $W_{RGPD}$ represent the Cramér–von Mises statistics, i.e.
		$W^2_{GPD}=\sum_{i=1}^{k}\left[\hat{F}_{GPD}\left(Z_i\right)-\left(2i-1\right)/\left(2n\right)\right]^2+1/\left(12n\right)
		$ and $W^2_{RGPD}=\sum_{i=1}^{k}\left[\hat{F}_{RGPD}\left(Z_i\right)-\left(2i-1\right)/\left(2n\right)\right]^2+1/\left(12n\right)$, where $\hat{F}_{GPD}$ and $\hat{F}_{RGPD}$ are the estimated cdf of the GPD and RGPD in our models. A small $W^2_{GPD}$ implies the estimated GPD fits well. The above procedure are summarized in Alogorithm 2.

		\begin{algorithm}
			\SetKwInOut{Input}{Input}
			\SetKwInOut{Output}{Output}
			\SetKwProg{Fn}{Algorithm}{}{\KwRet end}
			
			\caption{Confidence Interval for EVI Based on RGPD Perturbation}
			\Input{Data $\mathbf{Y} = \{Y_i\}_{i=1,2,\ldots,n}$, tail size $k$, perturbation scales $b_1, b_2$, confidence level $\alpha$}
			\Output{Confidence interval $I_{\text{Rptb}}$}
			
			\Fn{RptbPivotal$(\mathbf{Y}, k, b)$}{
				Calculate the Hill estimator $\hat{\gamma}$ using the order statistics $Y_{n-k,n} \le \ldots \le Y_{n,n}$\;
				
				Apply the RGPD likelihood maximization procedure to obtain $\hat{\btheta}=\left(\hat{\gamma}_{d}, \hat{\delta}, \hat{\tau}\right)$\;
				
				\For{$i = 1$ \KwTo $m$}{
					\For{$j = 1$ \KwTo $k$}{
						Generate $e_{ij}$ independently from Laplace$(0, b)$\;
						Generate $Z_{ij}^*$ by solving 
						\[
						G_{\hat{\btheta}}(Z_j)+ e_{ij} = H_{\hat{\gamma}, \hat{\delta}, \hat{\tau}}(Z_{ij}^*/ Y_{n-k,n}),
						\]
						where $H_{\gamma, \delta, \tau}(x)$ is the RGPD as defined in \cite{beirlant2009second}\;
					}
					Calculate the perturbed Hill estimator
					\[
					\hat{\gamma}_{i}^* = \frac{1}{k}\sum_{j=1}^{k} \log\left(\frac{Z_{ij}^* + Y_{n-k,n}}{Y_{n-k,n}}\right)
					\]
					Calculate the perturbed pivotal $T_i^* = T(\hat{\gamma}, \hat{\gamma}_{i}^*)$\;
				}
				\Return $\{T_i^*, i = 1, 2, \ldots, m\}$ and $\hat{\gamma}$\;
			}
			
			
			
			Compute $(\{T_{GPD,i}^*\}_{i=1}^{m} ,\hat{\gamma})\gets \textit{PtbPivotal}(\mathbf{Y}, k, b_1)$\;
			
			Compute $(\{T_{RGPD,i}^*\}_{i=1}^{m} , \hat{\gamma})\gets \textit{RptbPivotal}(\mathbf{Y}, k, b_2)$\;
			Compute $W_{GPD},W_{GPD}$, and $\omega$\;
			\For{$i = 1$ \KwTo $m$}{
				Calculate the weighted pivotal statistic $T_{w,i}^* \gets w T_{GPD,i}^* + (1 - w) T_{RGPD,i}^*$\;
			}
			
			$I_{\text{Rptb}} \gets$ \textit{CIPivotal}$(\{T^{*}_{w,i}\}_{i=1}^m, \hat{\gamma})$.
		\end{algorithm}

		\section{Simulations}
		\label{sec:simul}
		In this section, we perform two simulation studies to show the usefulness of the method obtained in Section \ref{sec:ptb} and the refined method for a more difficult case in Section \ref{subsec:RGPD} in practice. Additionally, a sensitivity analysis of the parameter of the perturbation sacle $c_1$ is conducted. Throughout the simulation study, we consider the Hill estimators for the heavy-tailed distribution, i.e. $\gamma>0$. In Study 1, we simulate observations from three different distributions: Pareto$(\alpha)$ distribution with cdf $F(x ;  \alpha)=1-\left(1/{x}\right)^\alpha$, $x>1$, $\alpha>0$; Fréchet$(\alpha)$ distribution with cdf 
		$F(x)=\exp \left\{-x^{-\alpha}\right\}$, $x>0$, where $\alpha>0$; Hall and Welsh (HW) distribution \citep{hall1984best} with cdf $F(x)=1-x^{-1 / \gamma}\left(1+x^{-1 /(2 \gamma)}\right) / 2,x>1$. For Pareto$(\alpha)$ distribution, $\gamma=1/\alpha$ and $\rho=-\infty$; for Fréchet$(\alpha)$ distribution, $\gamma=1/\alpha$ and $\rho=-1$; for HW distribution, $\gamma>0$ and $\rho=-1/2$. For each distribution, we set $\gamma=0.2$ and $0.5$. In Study 2, to verify the performence of the methods proposed in Section \ref{subsec:RGPD}, we generate data from the student $t$ distribution with the degree of freedom $\nu>0$, denoted as $t(\nu)$. Note that for $t(\nu)$, $\gamma=1/\nu$, and $\rho=-2/\nu$. Through the analysis in Section \ref{subsec:RGPD}, the small $\nu$ implies the large $|\rho|$ and thus the large asymptotic bias. We set $\nu=3$ and sample size $n=500,1000$ and $2000$. To make the choice of tail size $k$ comparable across different methods, we consider $k=\left[c_0 n^{1/3}\right]$ with $c_0\in\left[1,5\right]$ like \cite{wang2012estimation}. For each scenario, the simulation is repeated $B=500$ times.
		
		We apply the following methods to construct the $95\%$ confidence intervals for EVI: the proposed methods in Algorithm 1, denoted as Ptb; the proposed methods in Algorithm 2, denoted as RPtb; the parametric bootstrap estimator with percentile interval for asymptotic pivotal statistics, denoted as Para; the normal approximation methods in \cite{cheng2001confidence}, denoted as AN; the full-sample bootstrap estimator in \cite{de2024bootstrapping}, denoted as Boot.
		For Ptb and RPtb, we use the perturbation scale $b_1=2.5k^{-0.51}$ and $b_2=2*b_1$, respectively. For Boot, we use the $2.5\%$ and $97.5\%$ percentiles of bootstrap estimators to construct the confidence intervals. In Study 1, we compare the performance of Ptb, AN, Para and Boot. In Study 2, we compare the performence of RPtb, AN and Boot.
		
		For each scenario, we calculate the the coverage rate and the average length of the $B$ simulations. Here the coverage rate is defined as $ {B}^{-1}\sum_{i=1}^{B}\textbf{1}\left(\gamma\in I_i\right)$, where $I_i$ reprents the confidence interval in the $i$-th simulation. 
		
\begin{figure}
			\centering
			\subfigure[Pareto, $n=500$]{
				\begin{minipage}[b]{0.3\textwidth}
					\includegraphics[width=1\textwidth]{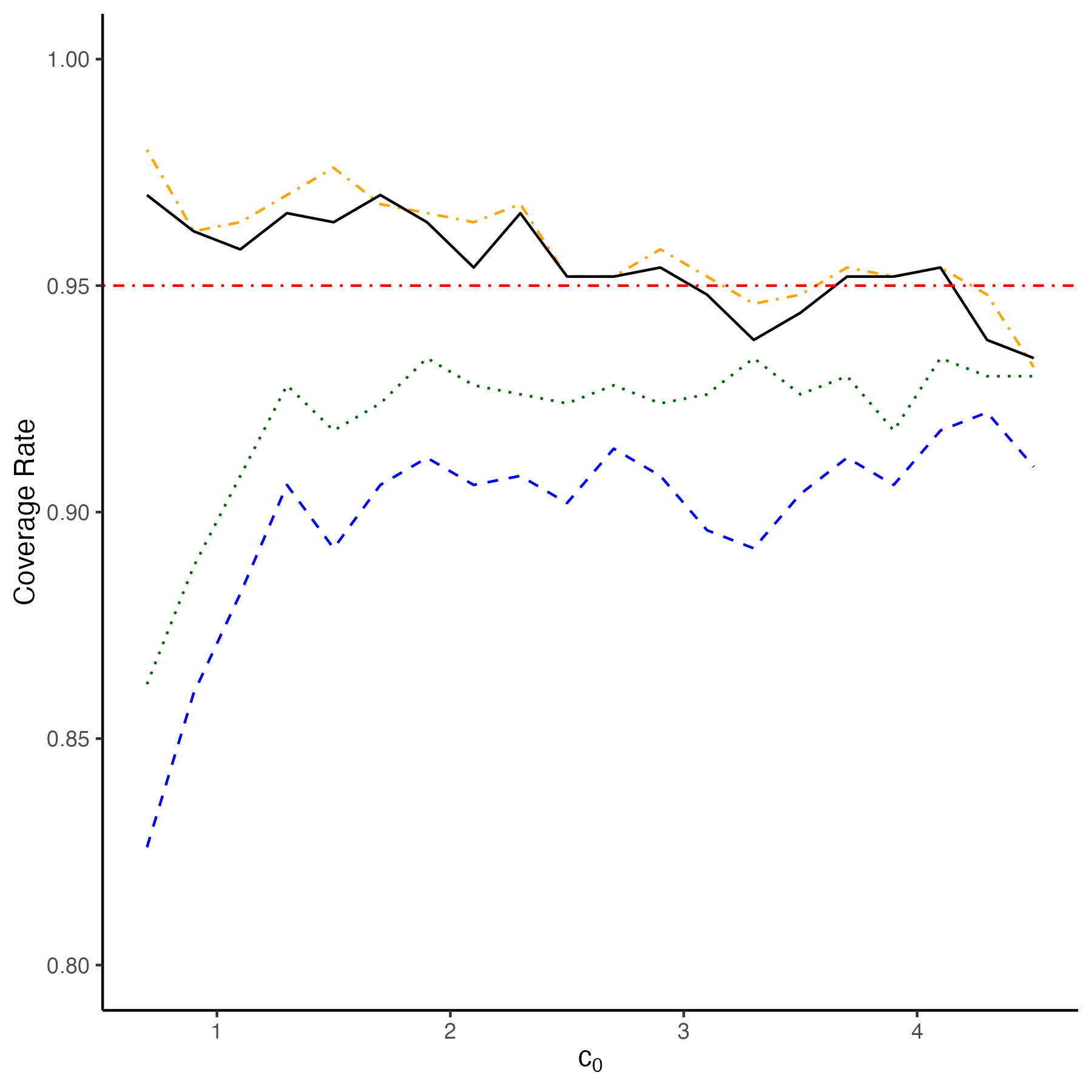}
				\end{minipage}
			}\subfigure[Fréchet, $n=500$]{
				\begin{minipage}[b]{0.3\textwidth}
					\includegraphics[width=1\textwidth]{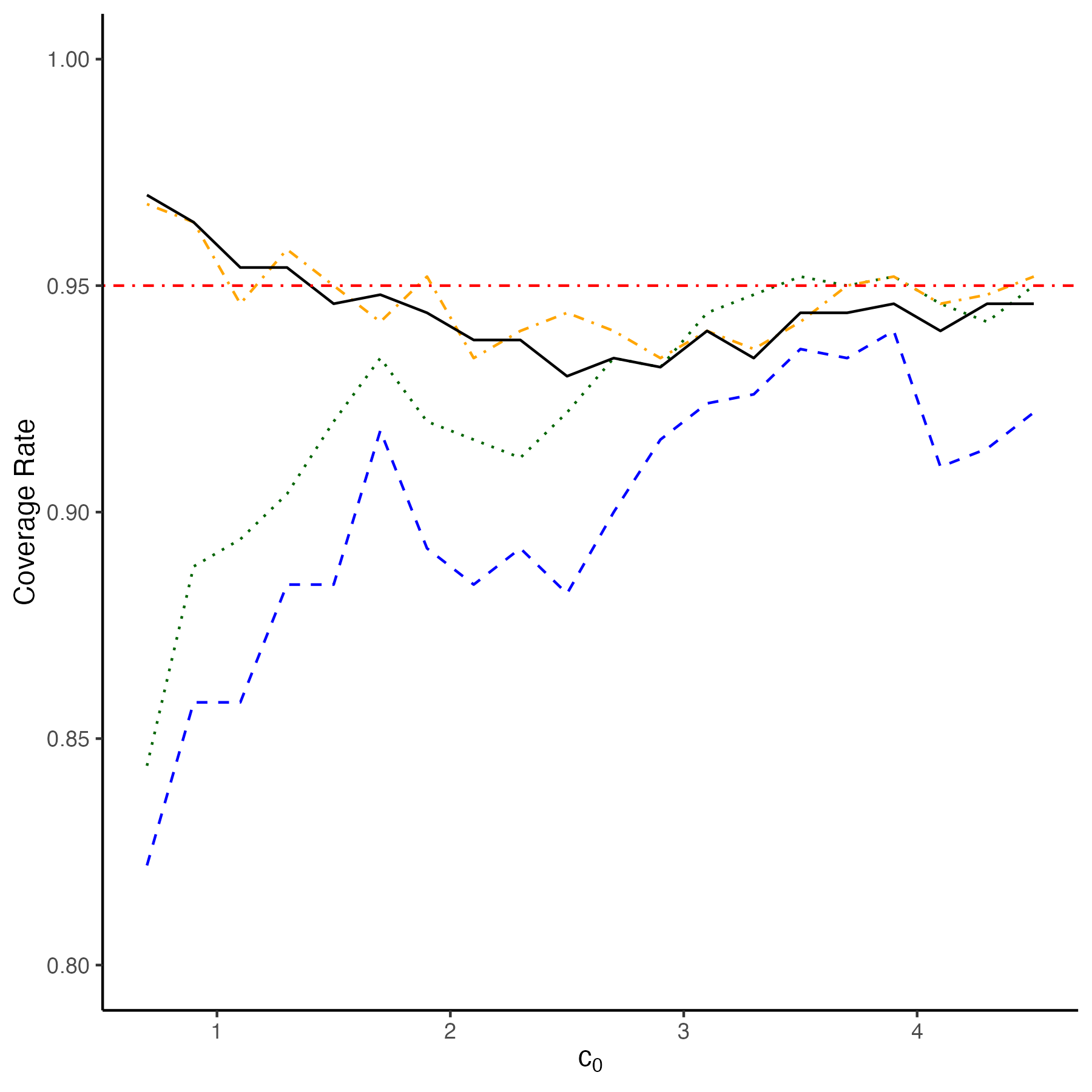}
				\end{minipage}
			}\subfigure[HW, $n=500$]{
				\begin{minipage}[b]{0.3\textwidth}
					\includegraphics[width=1\textwidth]{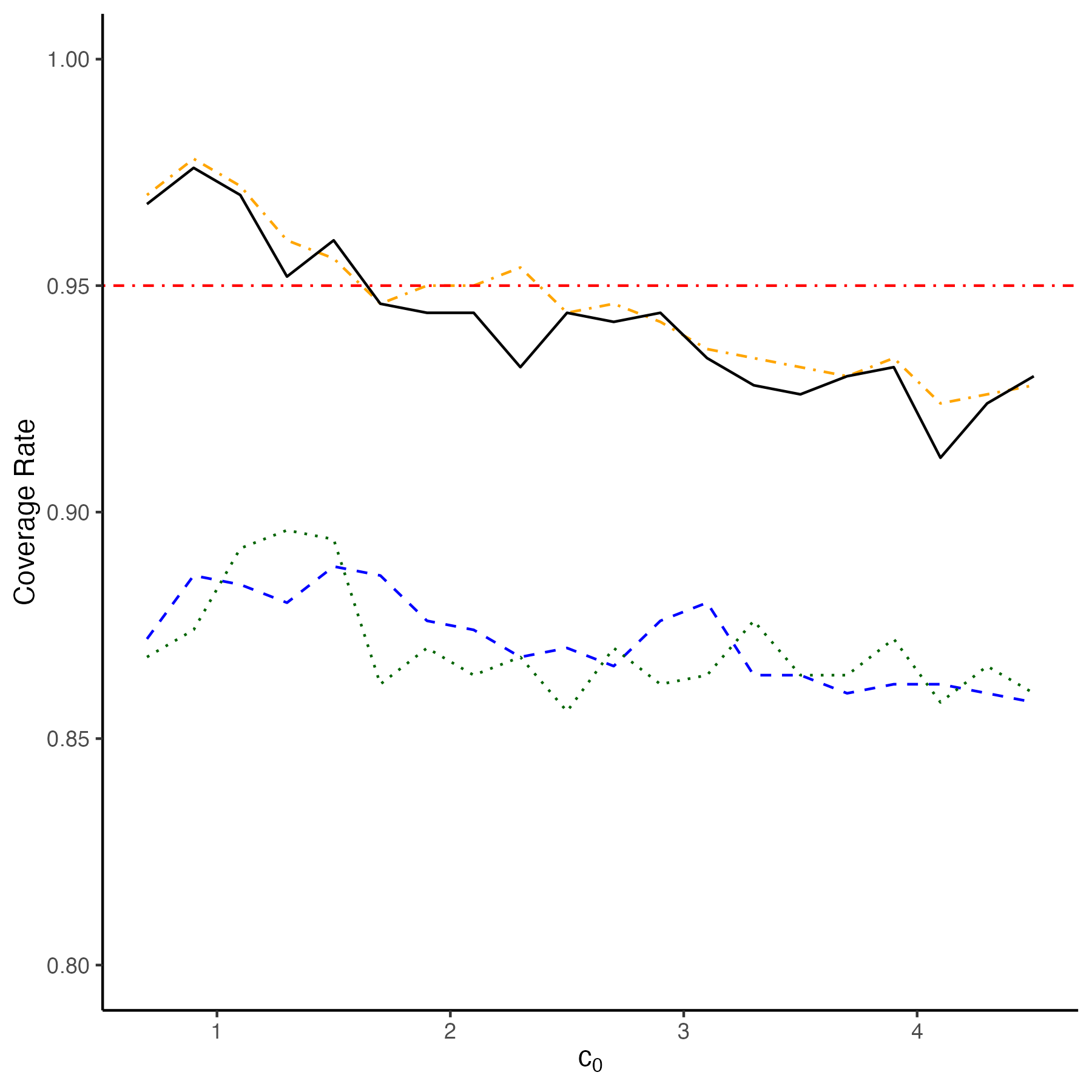}
				\end{minipage}
			}
			\subfigure[Pareto, $n=2000$]{
				\begin{minipage}[b]{0.3\textwidth}
					\includegraphics[width=1\textwidth]{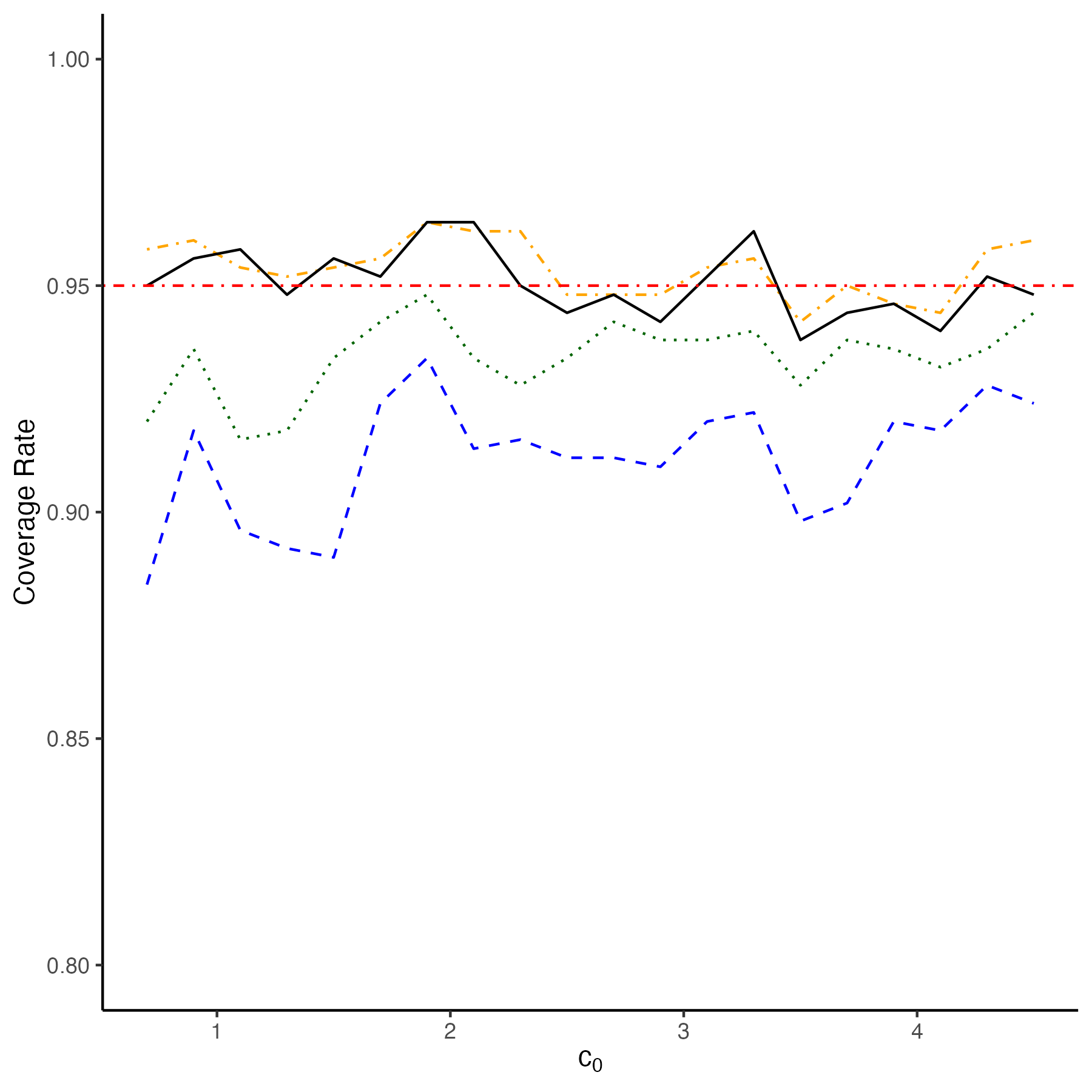}
				\end{minipage}
			}\subfigure[Fréchet, $n=2000$]{
				\begin{minipage}[b]{0.3\textwidth}
					\includegraphics[width=1\textwidth]{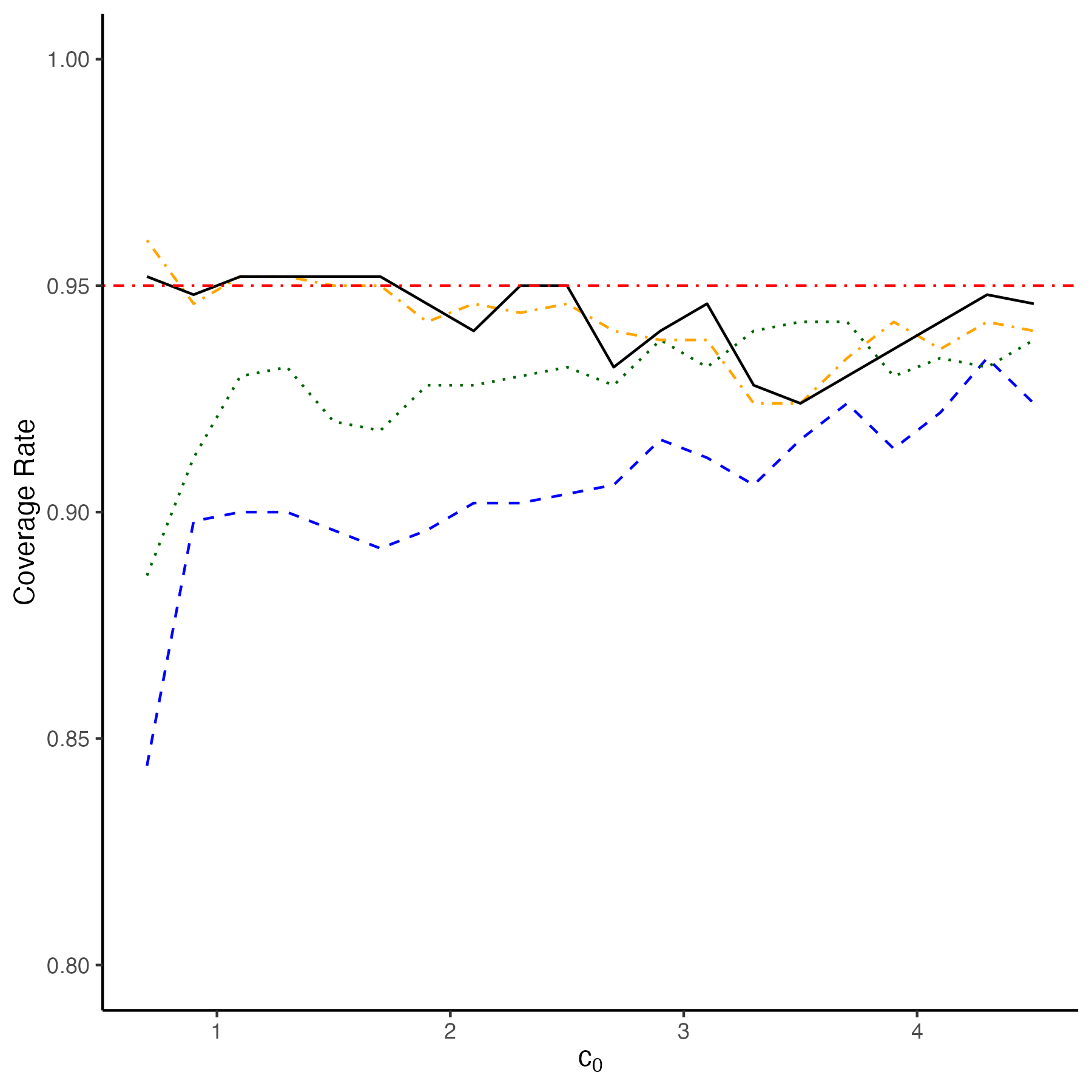}
				\end{minipage}
			}\subfigure[HW, $n=2000$]{
				\begin{minipage}[b]{0.3\textwidth}
					\includegraphics[width=1\textwidth]{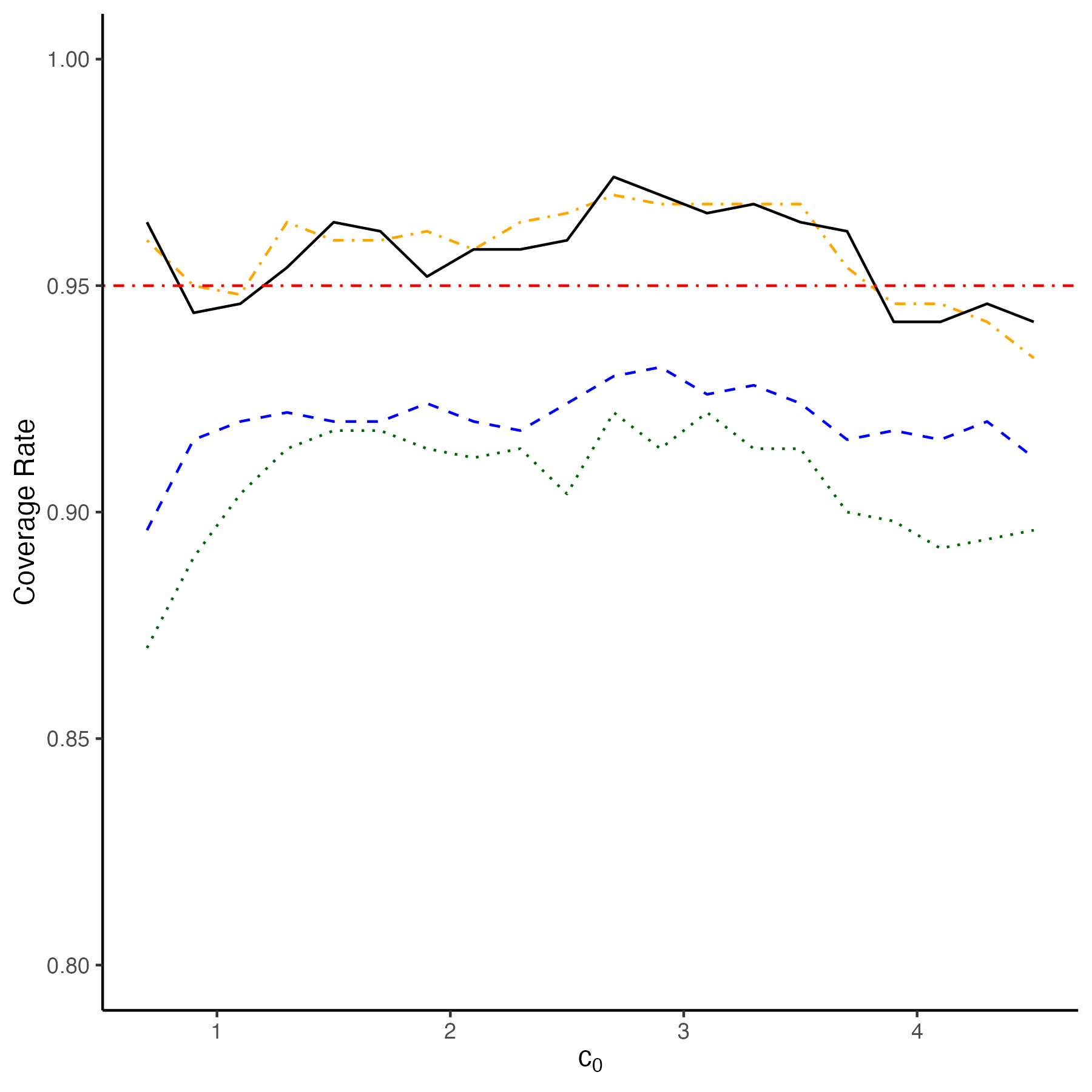}
				\end{minipage}
			}
			\caption{Covarage rates of confidence intervals for different methods with $\gamma=0.2$ and $n=500$ and $n=2000$: Ptb (black solid line), Para (orange dotdash line), AN (green dotted line), Boot (blue dashed line)}
			\label{fig:CR-study1n5002000}
		\end{figure}

\begin{figure}
			\centering
			\subfigure[$n=500$]{
				\begin{minipage}[b]{0.3\textwidth}
					\includegraphics[width=1\textwidth]{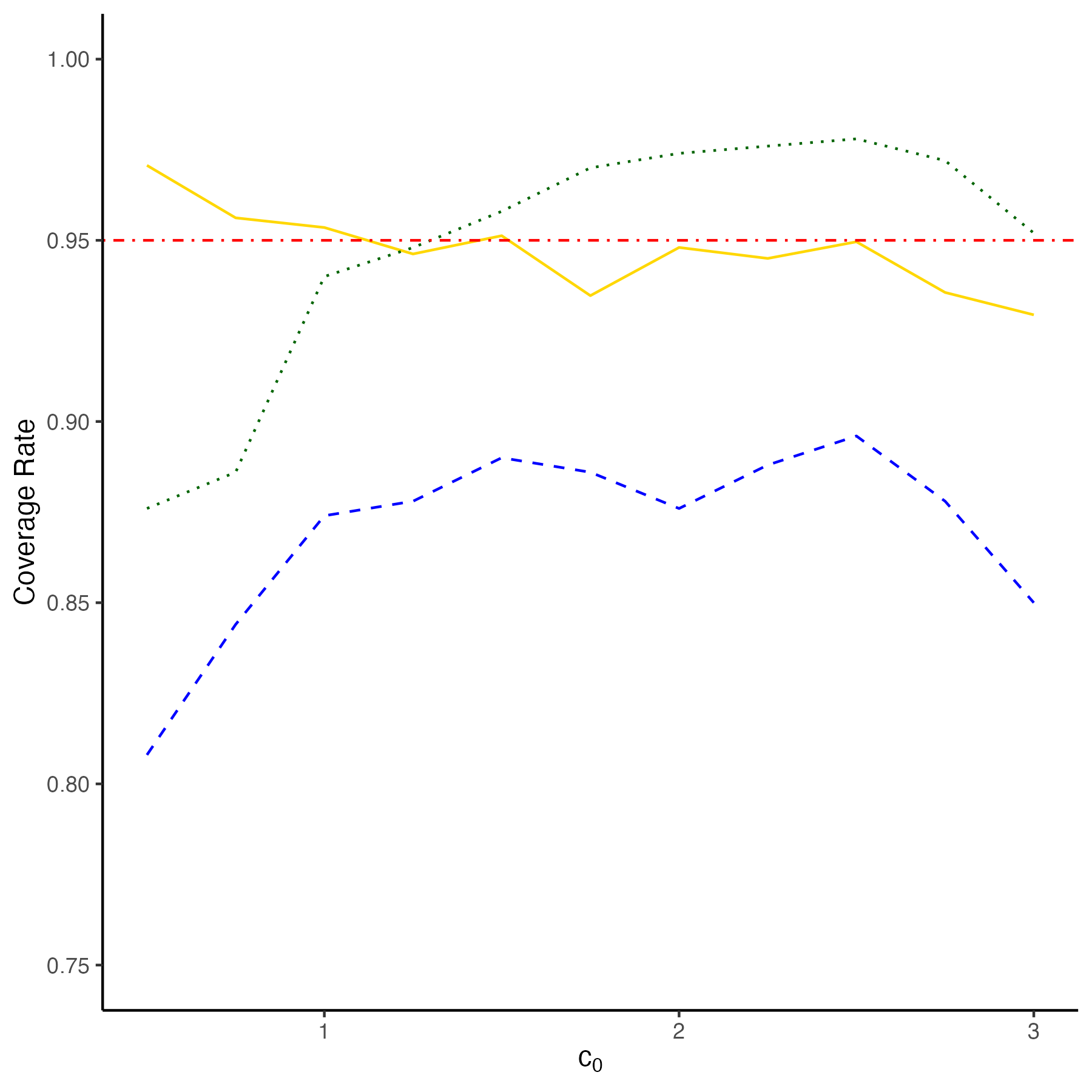}
				\end{minipage}
			}
			\subfigure[$n=1000$]{
				\begin{minipage}[b]{0.3\textwidth}
					\includegraphics[width=1\textwidth]{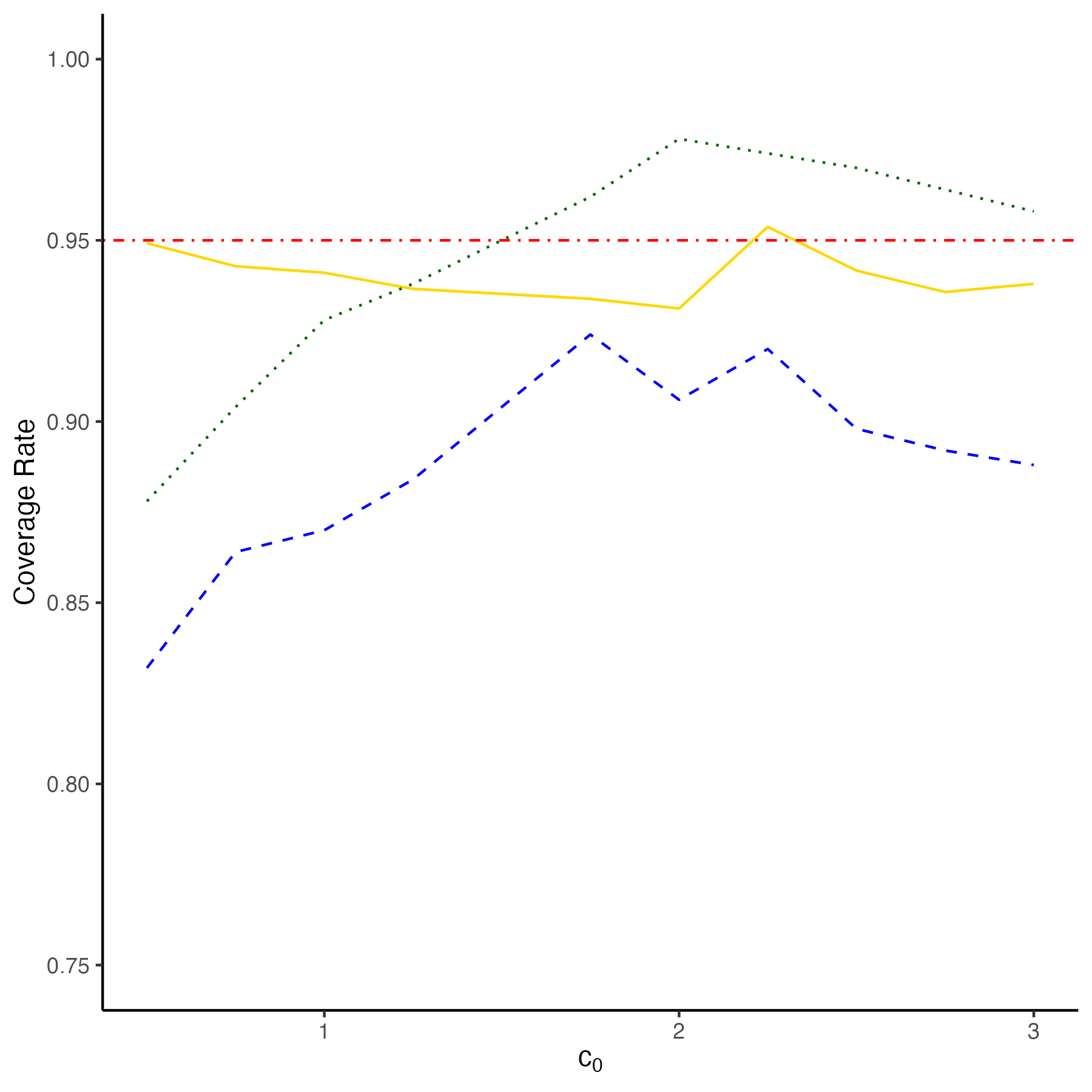}
				\end{minipage}
			}
			\subfigure[$n=2000$]{
				\begin{minipage}[b]{0.3\textwidth}
					\includegraphics[width=1\textwidth]{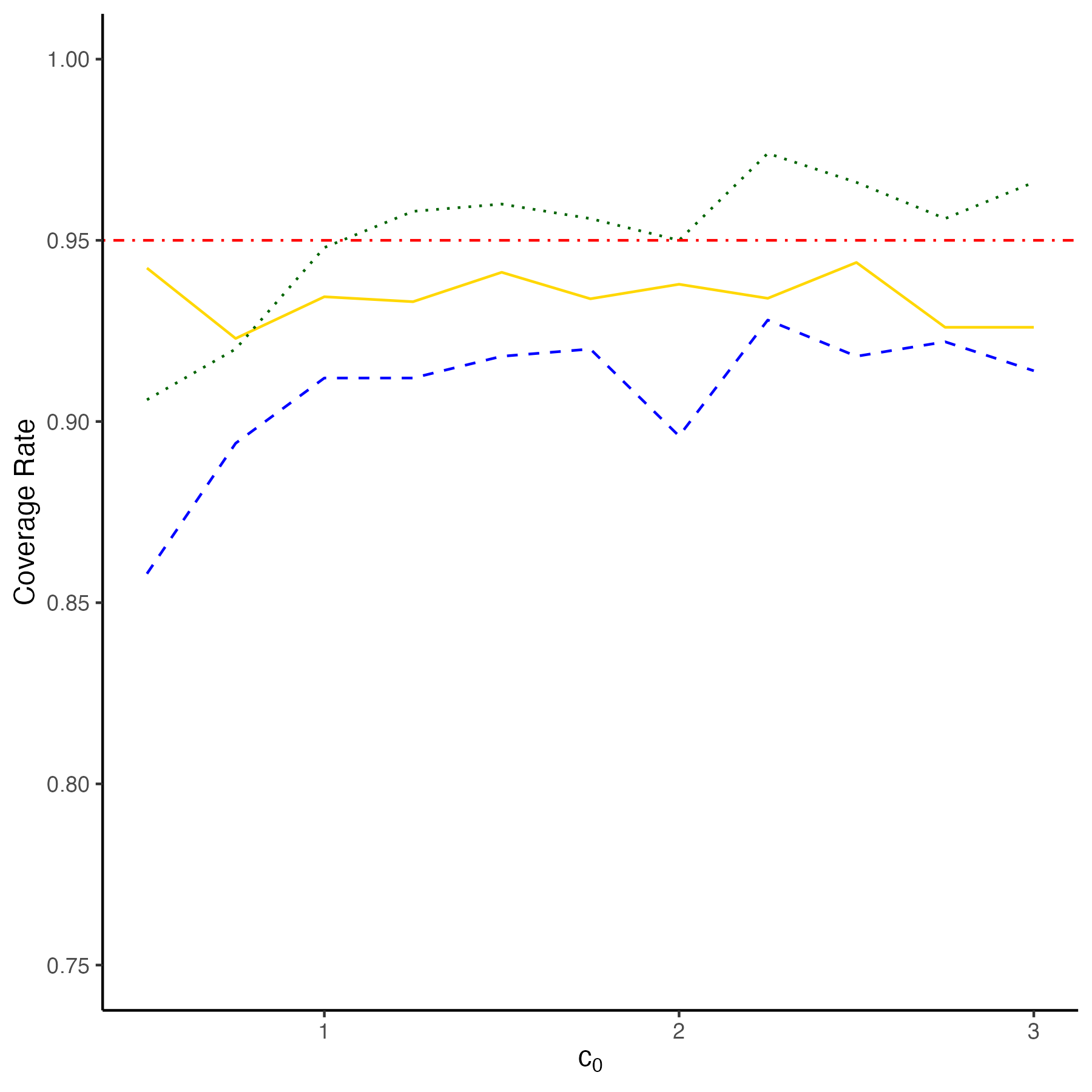}
				\end{minipage}
			}
			\caption{Covarage rates of confidence intervals for different methods under $t(3)$ distribution: RPtb (gold solid line), AN (green dotted line), Boot (blue dashed line)}
			\label{fig:CR-study2}
		\end{figure}

		We begin by comparing the coverage rate for the different confidence intervals. Figure \ref{fig:CR-study1n5002000} displays the coverage rates of different methods for different distributions with $\gamma=0.2$ and $n=500, 2000$, respectively. The results of $n=1000$ exhibit similar behavior and are presented in the supplementary material due to space limitation. For both sample sizes, the coverage rate of Ptb and Para remains stable with respect to $c_0$ and is approximately at $95\%$, indicating that Ptb and Para is nearly optimal for a $95\%$ confidence interval. For the smaller sample size of $n=500$, the coverage rates for AN and Boot are unstable with respect to $c_0$ and consistently fall below $95\%$, particularly for the HW distribution, which has a relatively large asymptotic bias. In contrast, for the larger sample size of $n=2000$, their coverage rates increase but still remain slightly below that of Ptb. Figure \ref{fig:CR-study2} illustrates the coverage rates of different methods for $t(3)$ distribution with $n=500, 1000, 2000$ from Study 2. Across all three sample sizes, the coverage rate of RPtb decreases gradually as $c_0$ increases but remains above $95\%$ when $c_0 < 2$. Meanwhile, the coverage rates of AN and Boot initially increase before sharply declining with respect to $c_0$, highlighting the importance of selecting an appropriate tail sample size $k$ in the literature on estimating extreme value index. The coverage rate of Boot is consistently below $95\%$, whereas AN maintains a coverage rate above $95\%$ only over a limited range of $c_0$.
		
\begin{figure}
			\centering
			\subfigure[Pareto, $n=500$]{
				\begin{minipage}[b]{0.3\textwidth}
					\includegraphics[width=1\textwidth]{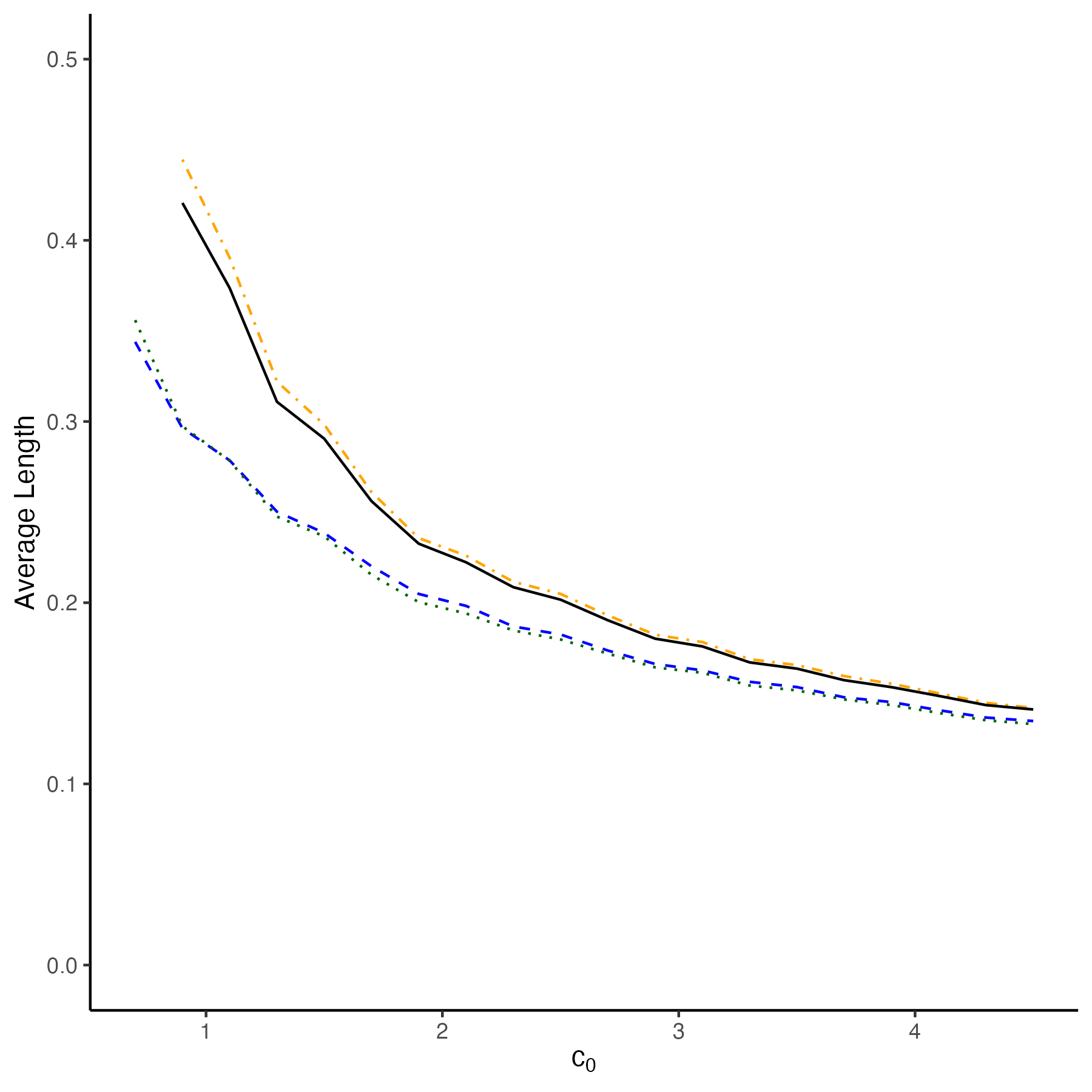}
				\end{minipage}
			}\subfigure[Fréchet, $n=500$]{
				\begin{minipage}[b]{0.3\textwidth}
					\includegraphics[width=1\textwidth]{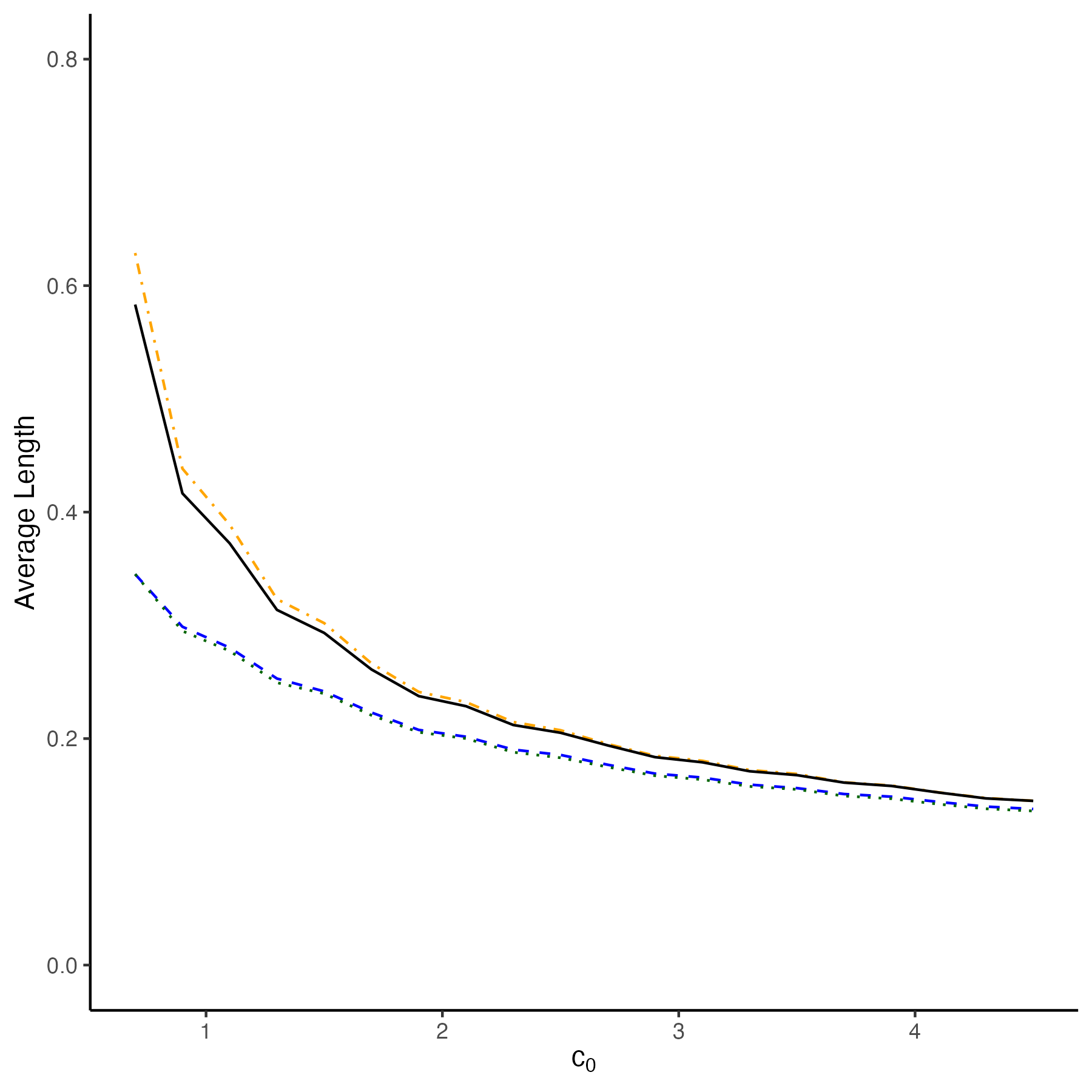}
				\end{minipage}
			}\subfigure[HW, $n=500$]{
				\begin{minipage}[b]{0.3\textwidth}
					\includegraphics[width=1\textwidth]{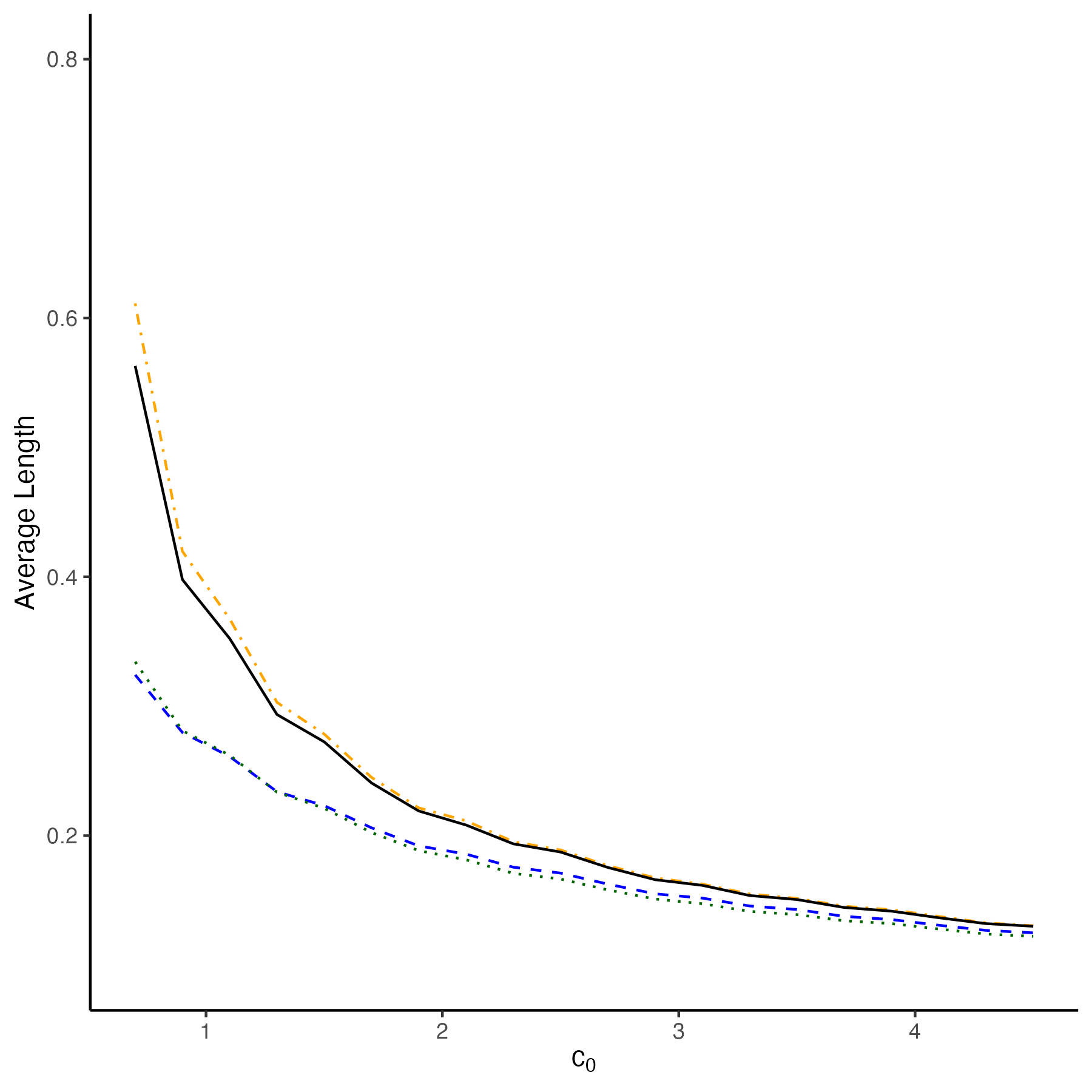}
				\end{minipage}
			}
			
			\subfigure[Pareto, $n=200$]{
				\begin{minipage}[b]{0.3\textwidth}
					\includegraphics[width=1\textwidth]{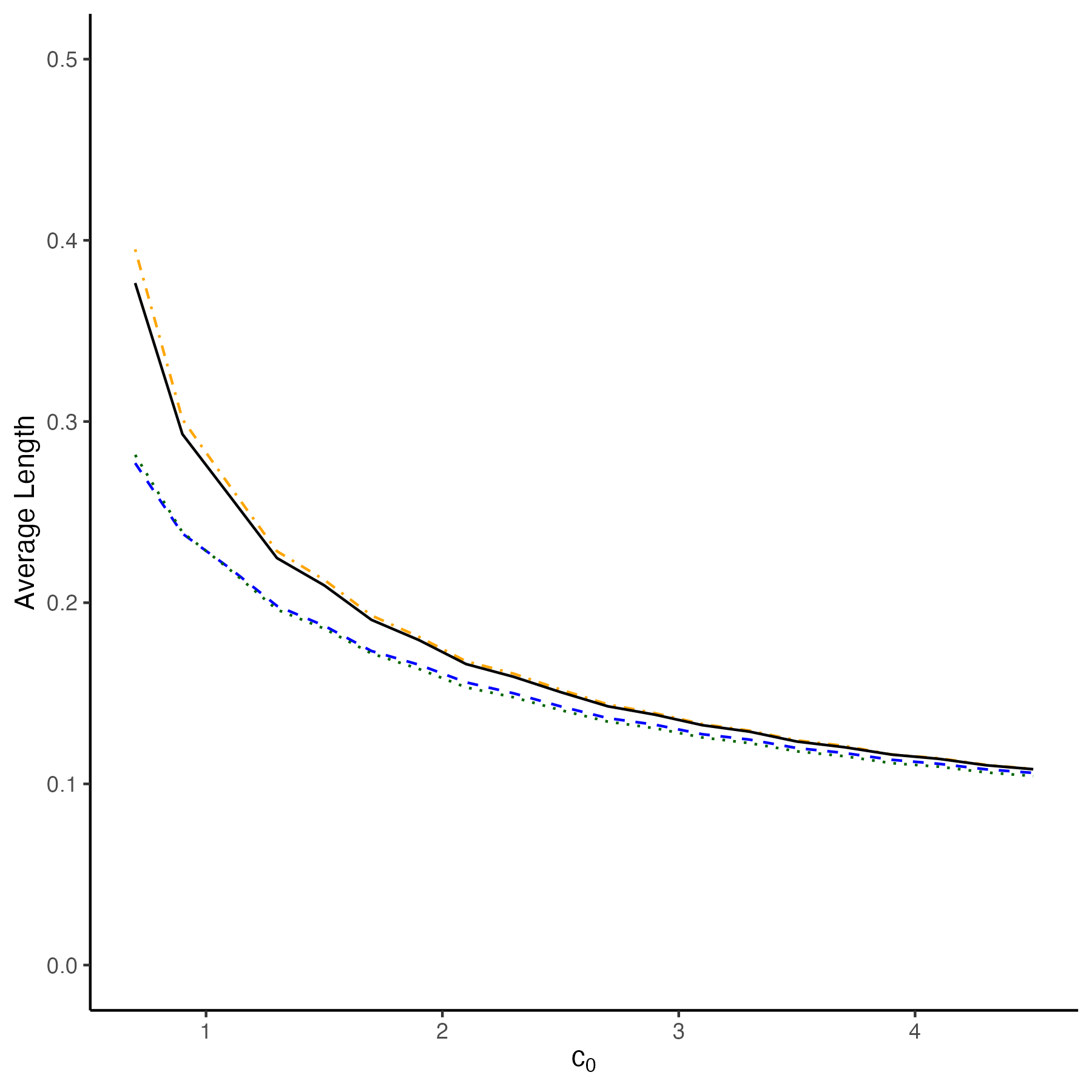}
				\end{minipage}
			}\subfigure[Fréchet, $n=200$]{
				\begin{minipage}[b]{0.3\textwidth}
					\includegraphics[width=1\textwidth]{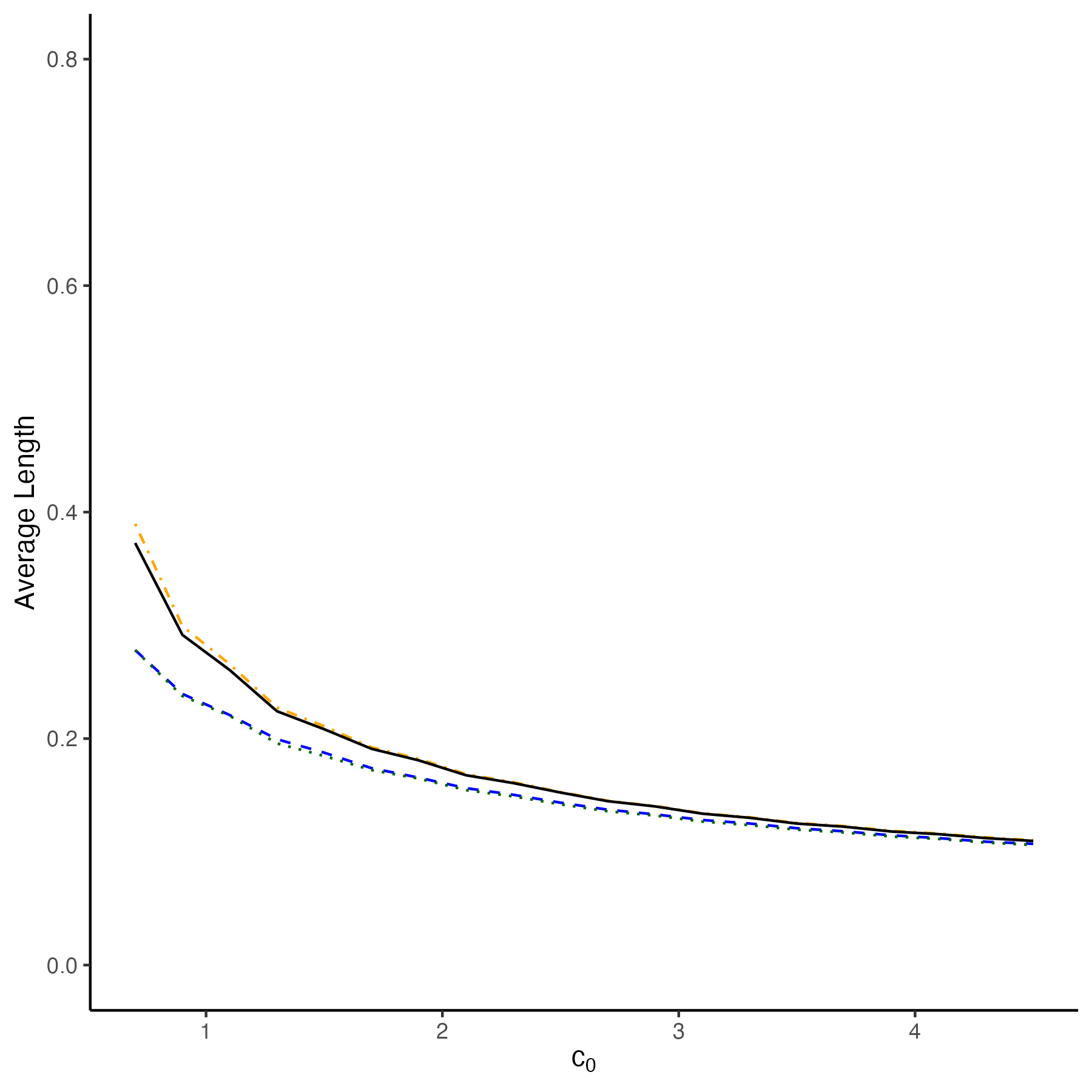}
				\end{minipage}
			}\subfigure[HW, $n=200$]{
				\begin{minipage}[b]{0.3\textwidth}
					\includegraphics[width=1\textwidth]{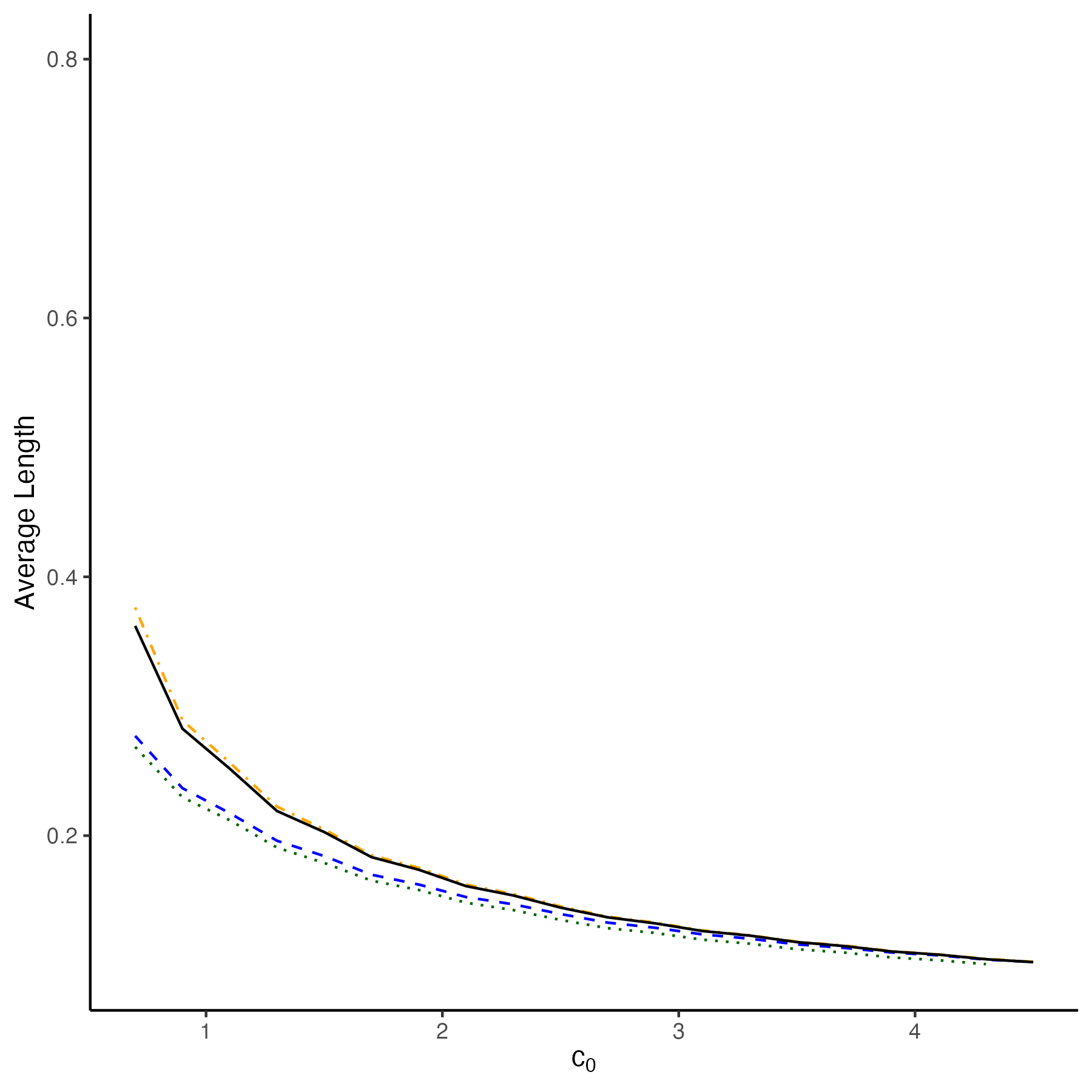}
				\end{minipage}
			}
			\caption{Average lengths of confidence intervals for different methods with $\gamma=0.2$ and $n=500$ and $n=2000$: Ptb (black solid line), Para (orange dotdash line), AN (green dotted line), Boot (blue dashed line)}
			\label{fig:AL-study1n5002000}
		\end{figure}

\begin{figure}
			\centering
			\subfigure[$n=500$]{
				\begin{minipage}[b]{0.3\textwidth}
					\includegraphics[width=1\textwidth]{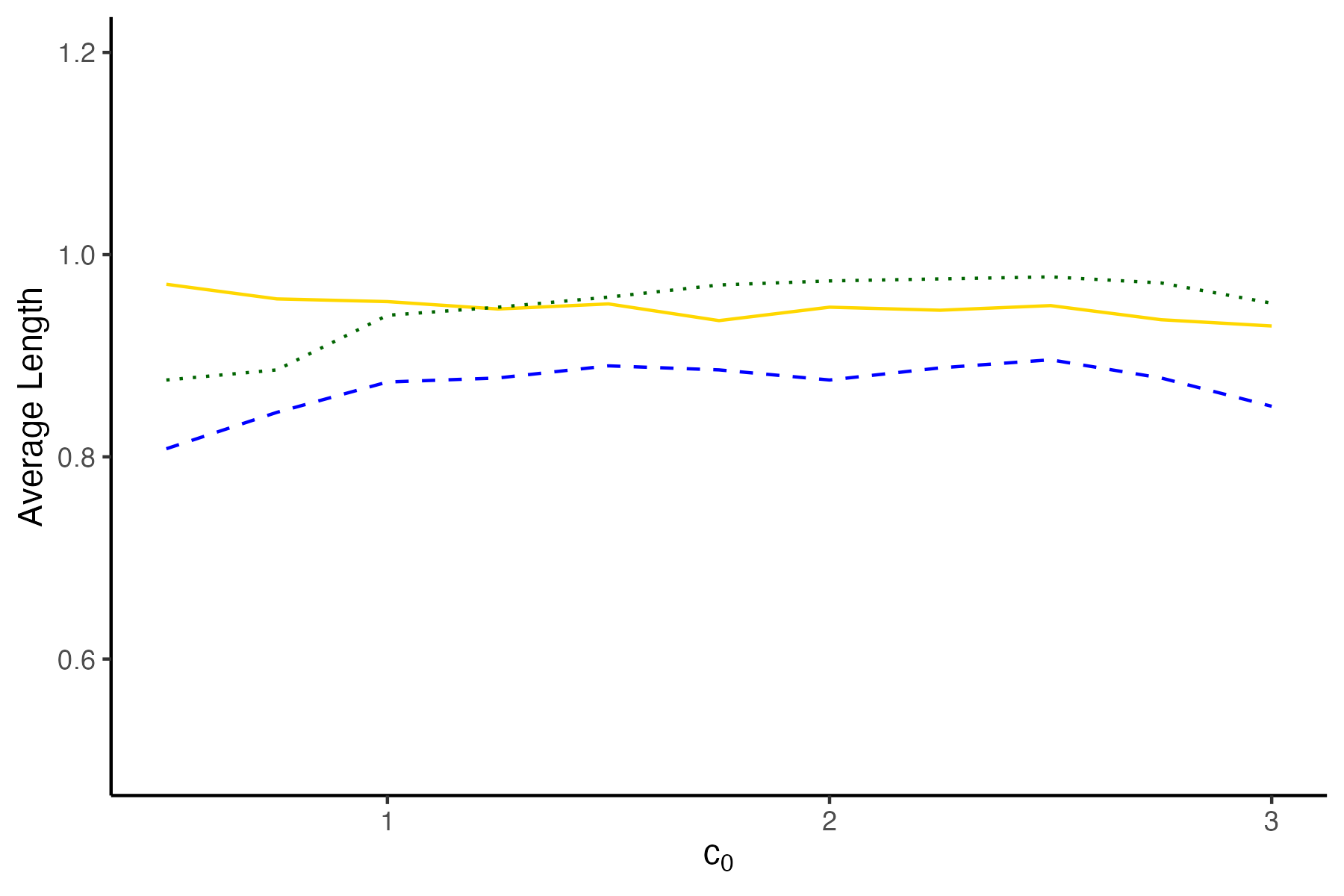}
				\end{minipage}
			}
			\subfigure[$n=1000$]{
				\begin{minipage}[b]{0.3\textwidth}
					\includegraphics[width=1\textwidth]{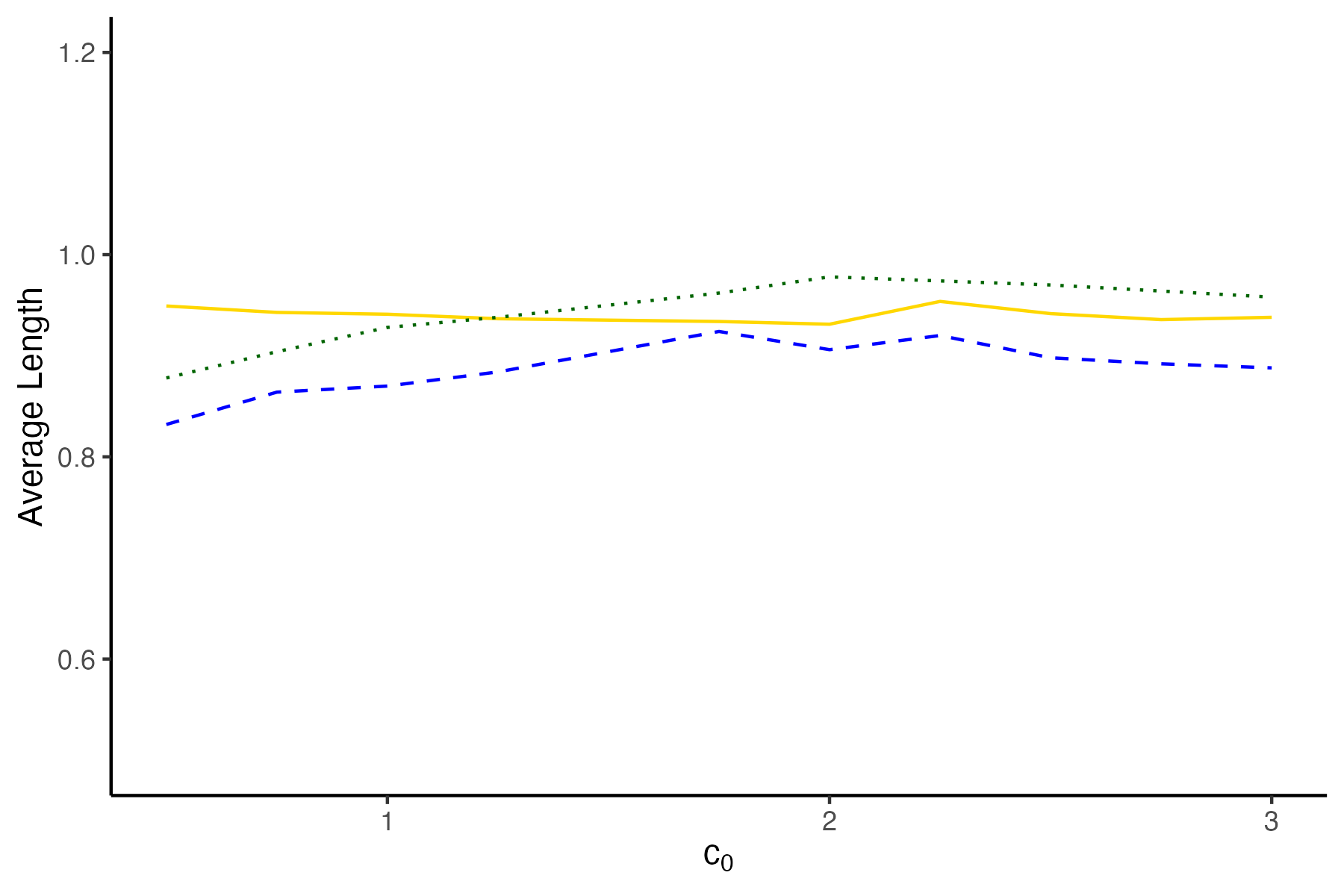}
				\end{minipage}
			}
			\subfigure[$n=2000$]{
				\begin{minipage}[b]{0.3\textwidth}
					\includegraphics[width=1\textwidth]{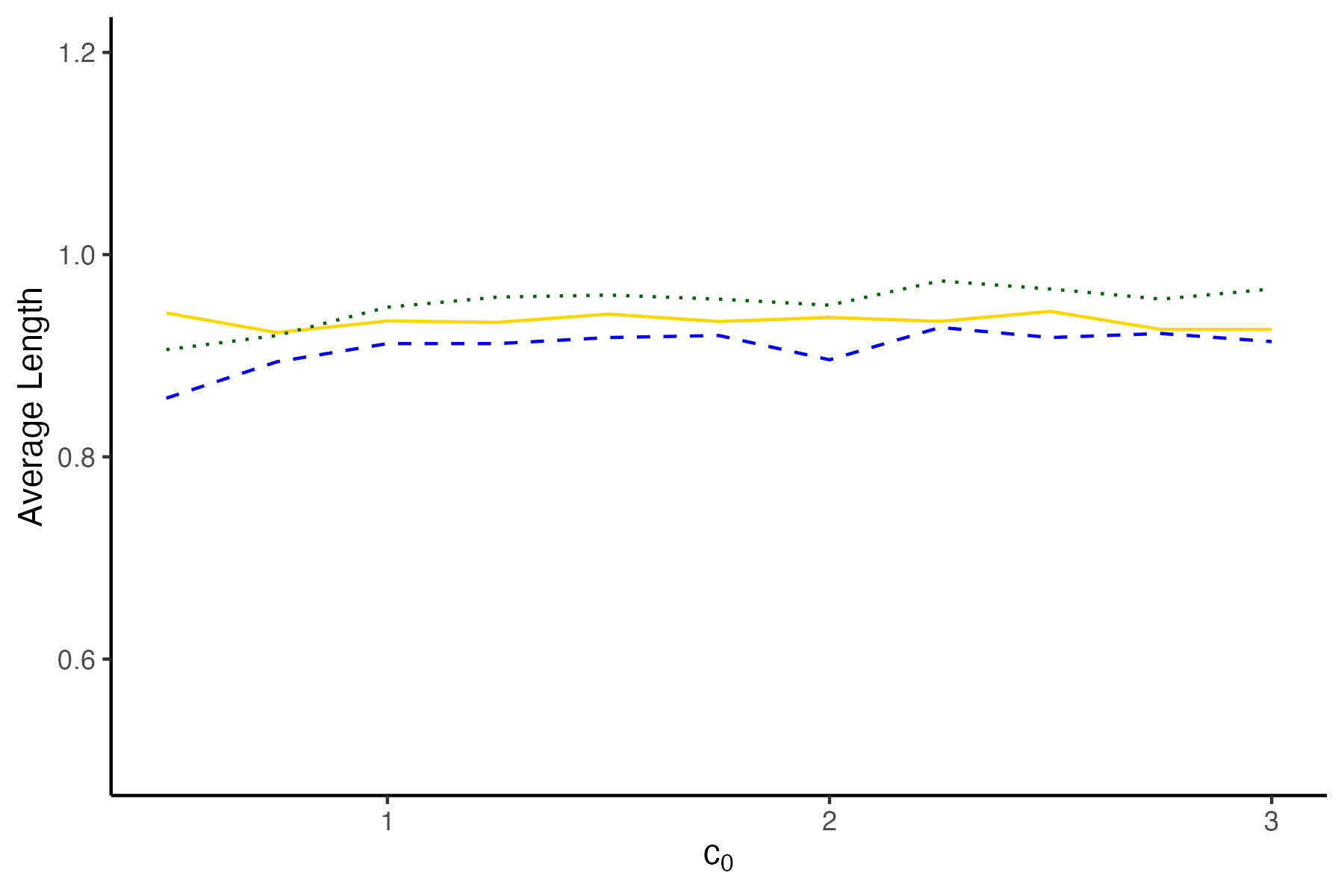}
				\end{minipage}
			}
			\caption{Average lengths of confidence intervals for different methods under $t(3)$ distribution: RPtb (gold solid line), AN (green dotted line), Boot (blue dashed line)}
			\label{fig:AL-study2}
		\end{figure}

		We next compare the average length of different confidence intervals. Figure \ref{fig:AL-study1n5002000} shows that as $c_0$ increases, the coverage rates of all four confidence intervals decrease and converge to a similar value. To achieve higher coverage rates, the average lengths of Ptb and Para are consistently and slightly longer than those of AN and Boot across these distributions, with Para a little bit longer. Additionally, the difference in average length diminishes when the sample size is large, such as $n=2000$.  Figure \ref{fig:AL-study2} shows the average lengths of different methods for the $t(3)$ distribution from Study 2. Here, the average length of Boot is consistently shorter than that of the other two confidence intervals, while the average lengths of RPtb and AN are comparable. The average length of RPtb remains relatively stable, whereas the average length of AN initially increases before decreasing as $c_0$ grows. Therefore, a relatively small tail sample size is recommended for RPtb.
		
		We perform a sensitivity analysis to assess the stability of the proposed procedure with respective to the perturbation scale in Ptb. For brevity, we only present results for the Pareto$(5)$ distribution; the other scenarios exhibit similar behavior, and the results are available in the Supplementary Material. Figures \ref{fig:sensitivity-CR-c1} shows the coverage rates and average lengths for Pareto$(5)$ against $c_1\in[1.5,3.5]$, $c_0\in \{1, 3, 5\}$, and \(k = [c_0 n^{1/3}]\) for \(n = 500, 1000, 2000\). The results indicate that both the average length and coverage rate of the confidence intervals show minimal variation with changes in the perturbation scale, particularly when \(c_0\) is small. As shown in Figures, coverage rate and average length is stable with respective to different $c_1$, which motivates us to choose $c_1=2.5$. Moreover, similar studies conducted for RPtb suggest that $c_1=2.5$ is preferable for that context.

		\begin{figure}
			\centering
			\subfigure[$n=500$]{
				\begin{minipage}[b]{0.28\textwidth}
					\includegraphics[width=0.9\textwidth]{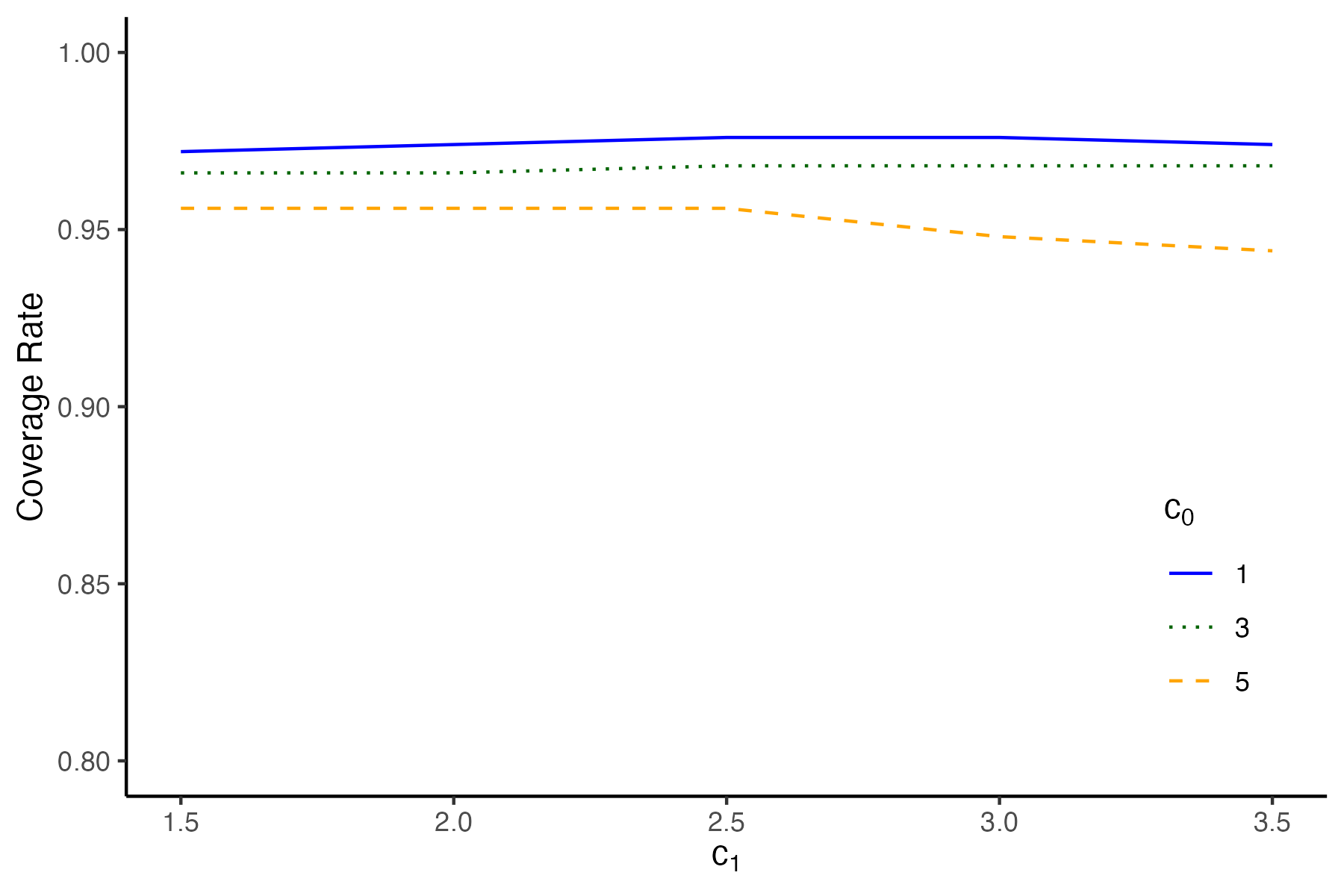}
				\end{minipage}
				\label{fig:sensitivity-CRsub1}
			}\subfigure[$n=1000$]{
				\begin{minipage}[b]{0.28\textwidth}
					\includegraphics[width=0.9\textwidth]{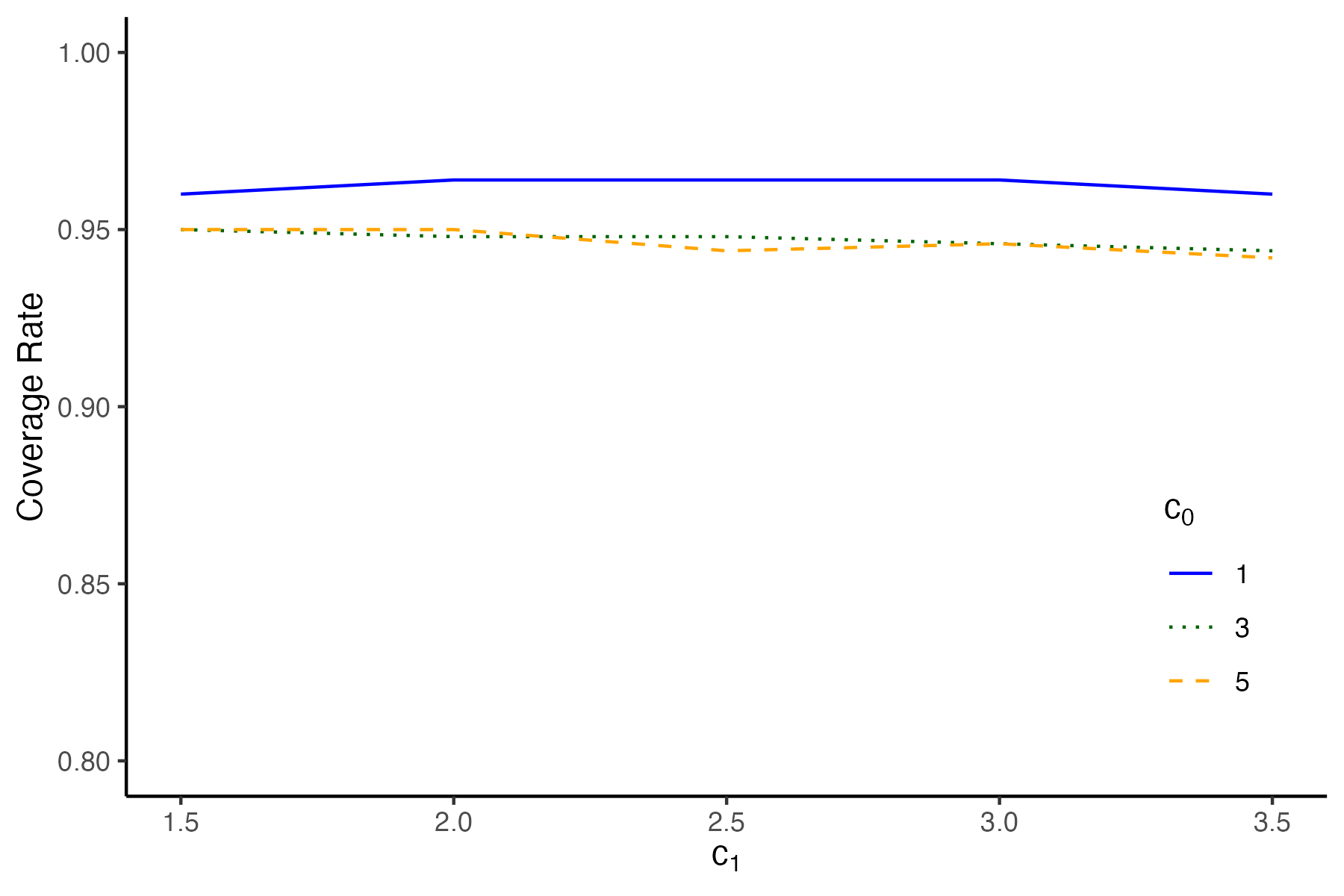}
				\end{minipage}
				\label{fig:tivity-CRsub2}
			}\subfigure[$n=2000$]{
				\begin{minipage}[b]{0.28\textwidth}
					\includegraphics[width=0.9\textwidth]{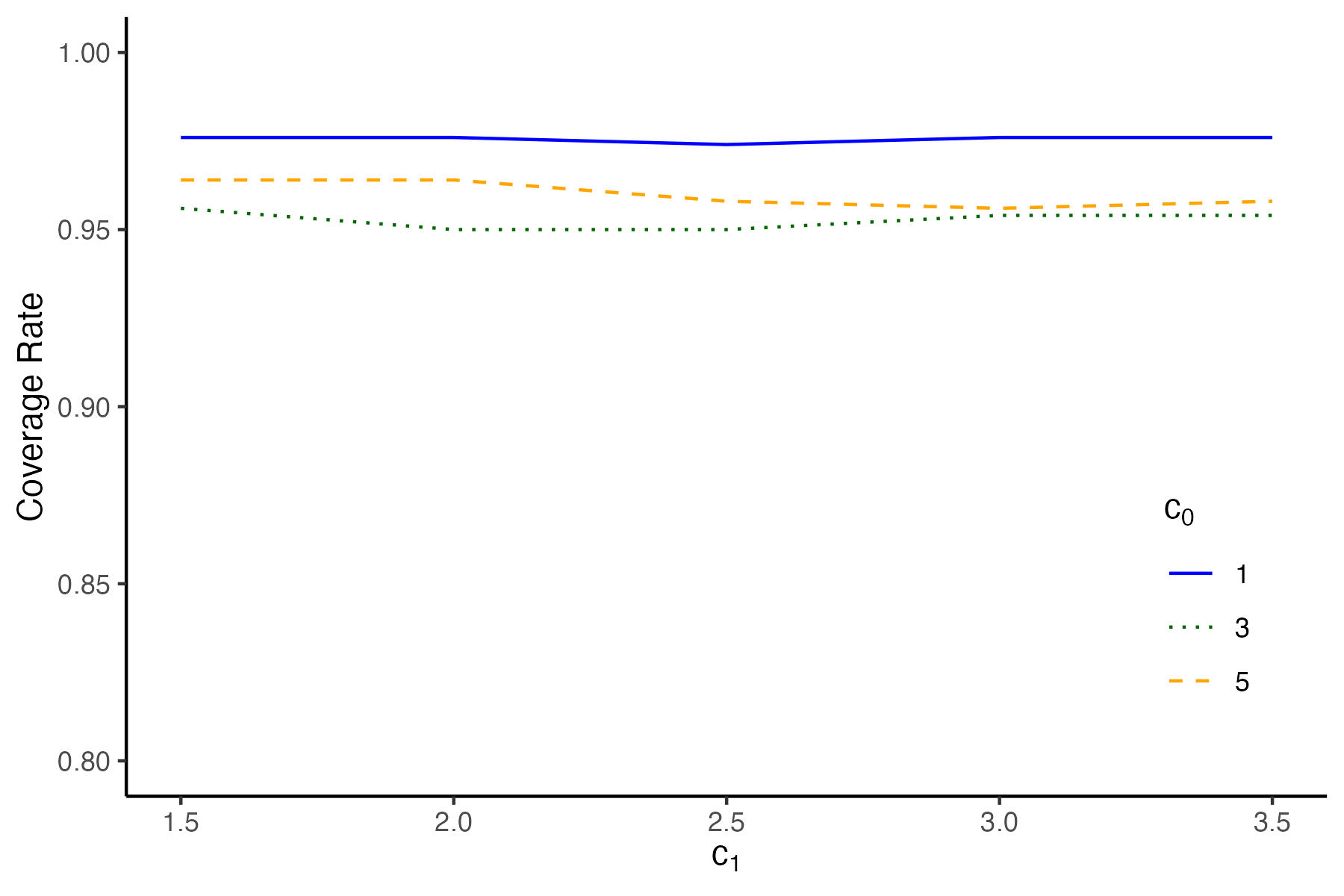}
				\end{minipage}
				\label{fig:sensitivity-CRsub3}
			}
			\subfigure[$n=500$]{
				\begin{minipage}[b]{0.28\textwidth}
					\includegraphics[width=0.9\textwidth]{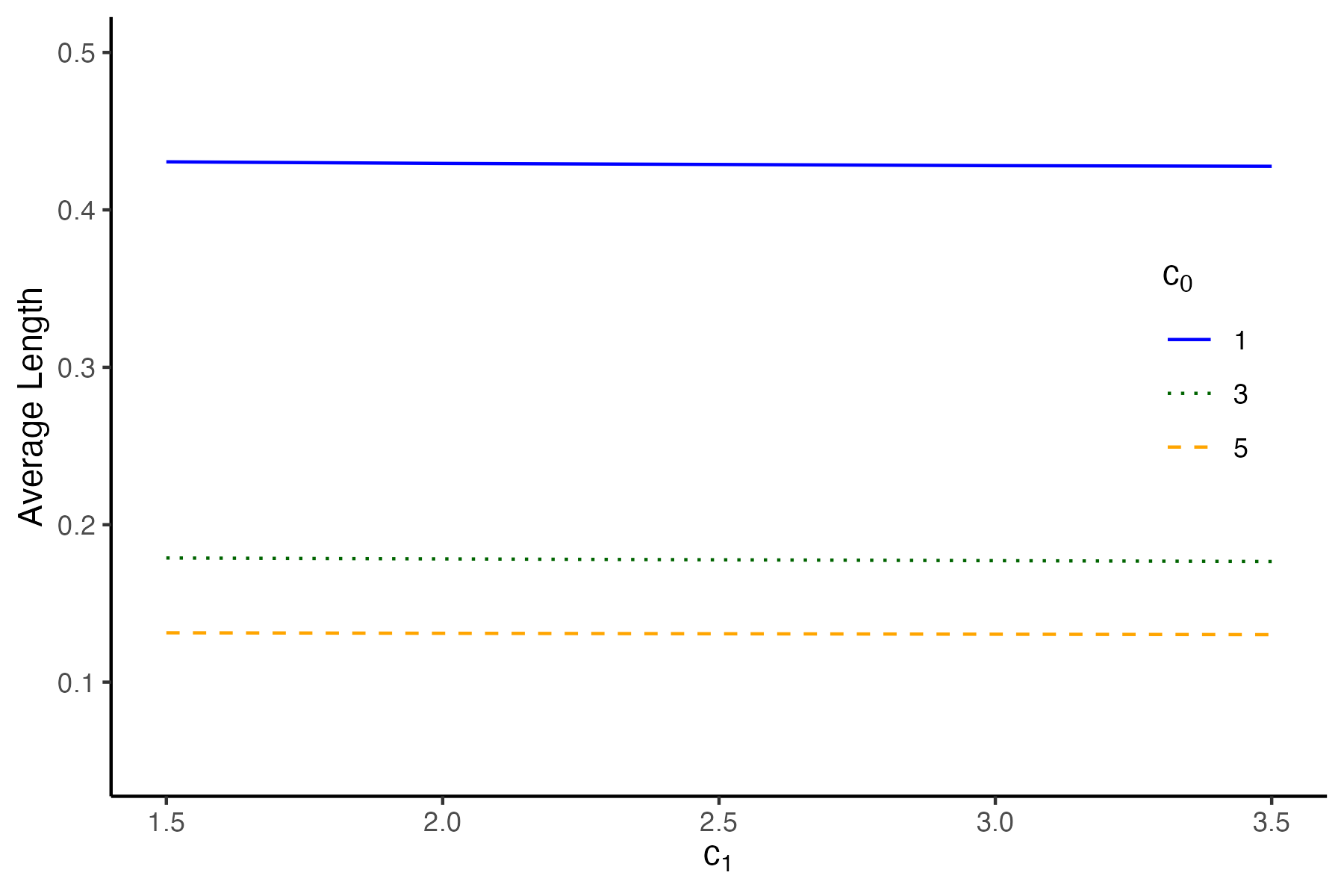}
				\end{minipage}
			}\subfigure[$n=1000$]{
				\begin{minipage}[b]{0.28\textwidth}
					\includegraphics[width=0.9\textwidth]{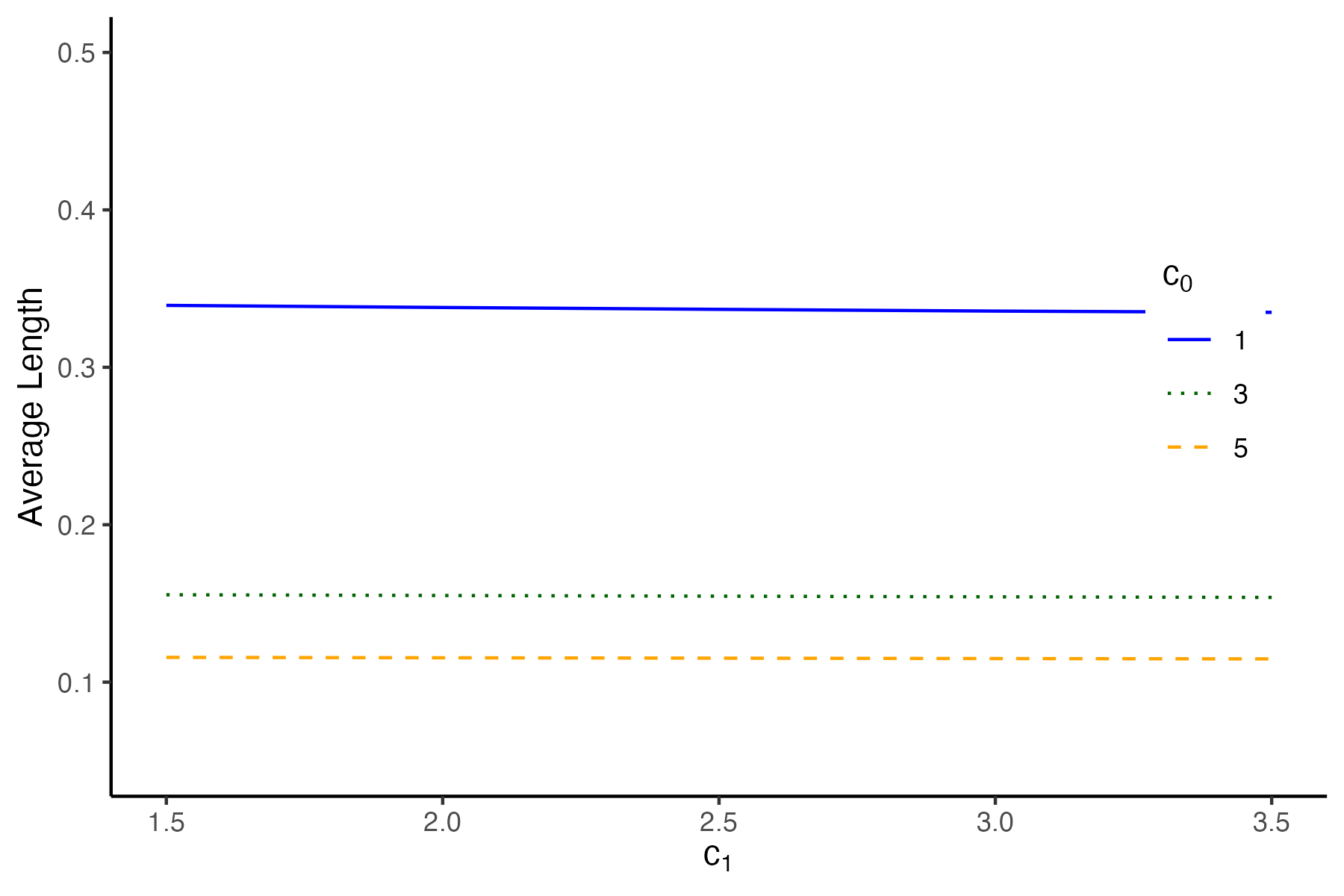}
				\end{minipage}
			}\subfigure[$n=2000$]{
				\begin{minipage}[b]{0.28\textwidth}
					\includegraphics[width=0.9\textwidth]{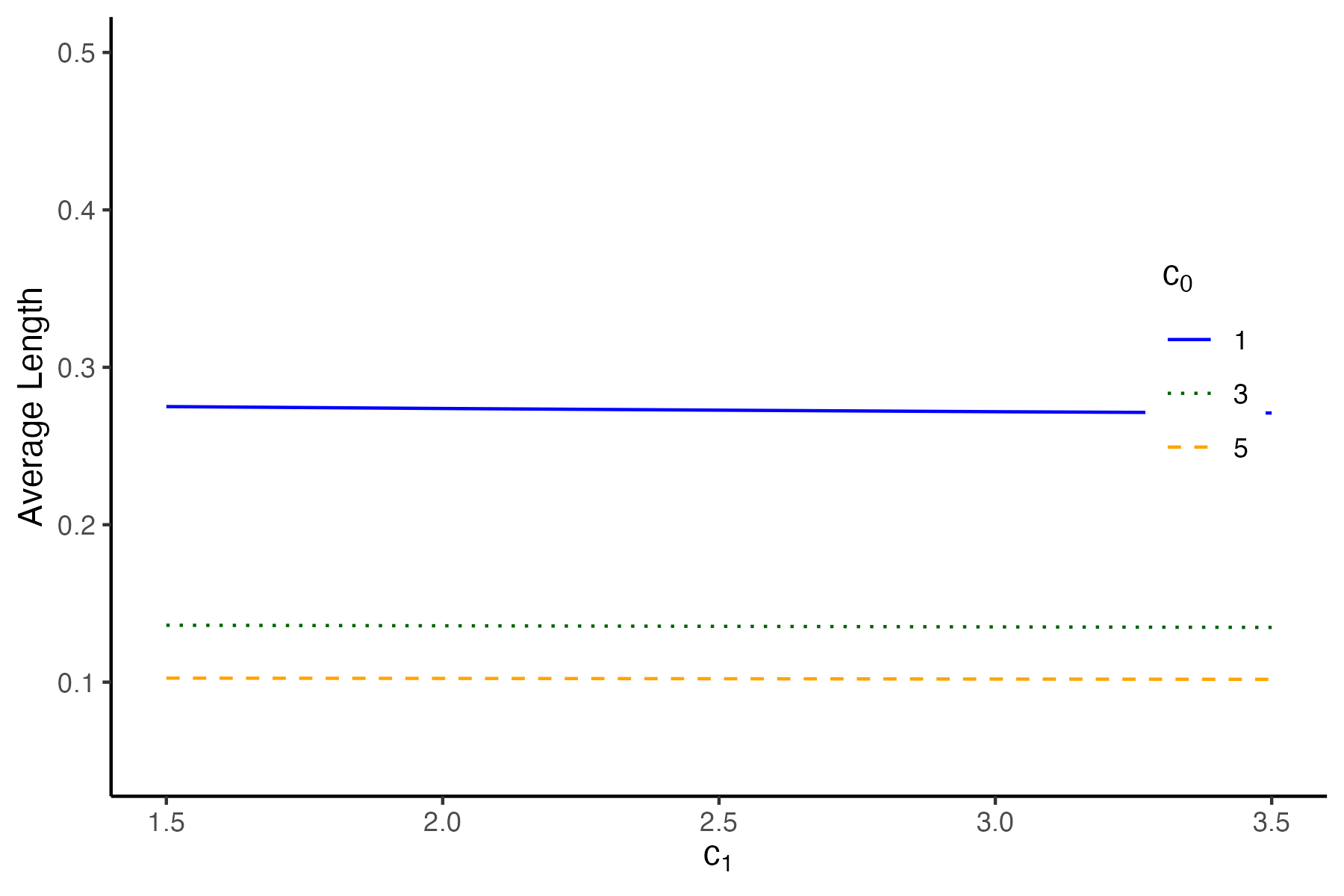}
				\end{minipage}
			}
			\caption{Sensitivity of coverage rates and average interval lengths to $c_1$ under the Pareto model with $\gamma = 0.2$, $k = [c_0 n^{1/3}]$, $c_0 \in \{1,3,5\}$, and $n=500$, $1000$, and $2000$.}
			\label{fig:sensitivity-CR-c1}
		\end{figure}

		\section*{Supplementary Materials}
		
		The supplementary materials are provided in a separate PDF entitled \textit{Supplementary Material for Perturbation-based Inference for Extreme Value Index}, which contains complete proofs of the theoretical results, additional discussions of the proposed methodology, and supplementary simulation results.
		
		\bibliographystyle{agsm1}
		\bibliography{literature_DataFlush}
	\end{document}